\documentclass[useAMS,usenatbib]{mn2e} 
\pdfoutput=1
\usepackage{journals}
\usepackage[pdftex]{graphicx,color} 
\usepackage[latin1]{inputenc}
\usepackage{graphics} 
\usepackage{amsfonts} 
\usepackage{amsmath}
\usepackage{multicol} 
\usepackage{layout} 
\usepackage{amssymb}
\usepackage[a4paper,colorlinks=true,pdfstartview=FitV,
linkcolor=red,citecolor=blue,urlcolor=magenta]{hyperref}
\title[Ellipsoidal haloes and models of triaxial halo formation]
{Ellipsoidal halo finders and implications for models of triaxial halo formation}
\author[Despali et al. 2012]
{\parbox{\textwidth}{Giulia Despali$^{1}$\thanks{E-mail:
 \href{mailto:giulia.despali@studenti.unipd.it} {giulia.despali@studenti.unipd.it}},
 Giuseppe Tormen$^{1}$, Ravi K. Sheth$^{2,3}$} \\ \\ 
 $^{1}$ Dipartimento di Fisica e Astronomia, Universit\`a di Padova, vicolo
dell'Osservatorio 3, 35122, Padova, Italy \\ 
 $^{2}$ The Abdus Salam International Center for Theoretical Physics, 
Strada Costiera 11, 34151 Trieste, Italy \\
 $^{3}$ Center for Particle Cosmology, University of Pennsylvania, 
       209 S 33rd St, Philadelphia, PA 19104, USA}
\begin{document}
\date{}
\maketitle
\label{firstpage}
\pagerange{\pageref{firstpage}--\pageref{lastpage}} \pubyear{2012}
\begin{abstract}
We  describe  an algorithm  for  identifying  ellipsoidal haloes  in
  numerical simulations,  and quantify how the  resulting estimates of
  halo mass and  shape differ with respect to  spherical halo finders.
  Haloes become  more prolate  when  fit  with  ellipsoids, the
  difference being  most pronounced  for the more aspherical objects.
  Although the  ellipsoidal mass is  systematically larger, this
  is less  than 10\% for  most of the  haloes.  However,
  even this small difference in mass corresponds to a significant
  difference in shape.   We quantify  these effects  also on the  initial  mass and
  deformation tensors,  on which most models of  triaxial collapse are
  based.

  By studying the properties of protohaloes in the initial conditions,
  we  find  that  models   in  which  protohaloes  are  identified  in
  Lagrangian space  by three  positive eigenvalues of  the deformation
  tensor  are  tenable  only  at  the masses  well-above  $M_*$.   The
  overdensity $\delta$ within almost  any protohalo is larger than the
  critical  value associated  with spherical  collapse  (increasing as
  mass decreases);  this is in good qualitative  agreement with models
  which identify haloes requiring that collapse have occured along all
  three  principal  axes, each  axis  having  turned  around from  the
  universal  expansion  at a  different  time.   The distributions  of
  initial  values  are  in  agreement with  the  simplest  predictions
  associated with  ellipsoidal collapse, assuming  initially spherical
  protohaloes,  collapsed around  random  positions which  were
  sufficiently overdense.

  However,  most protohaloes  are  not spherical  and departures  from
  sphericity  increase  as protohalo  mass  decreases.   The mass  and
  deformation  tensors   are  well-aligned,  in   agreement  with  the
  fundamental assumption  of ellipsoidal collapse, and with
  models  which identify  haloes  with peaks  in  the initial  density
  fluctuation field.  But the direction of maximum initial compression
  coincides with the direction of  what is initially the longest axis,
  contrary to what the peaks model predicts.  By the final time, it is
  the shortest axis of the final object which tends to be aligned with
  the direction of initial maximal compression: the alignment changes
  during the evolution.  
\end{abstract}

\section{Introduction}

In the standard cosmological  model, the structures observed today are
assumed to  have grown gravitationally from  small, initially Gaussian
density fluctuations.   Collapsed, virialized dark  matter haloes form
from the initial fluctuation field, leading to structure formation: as
the universe expands, sufficiently overdense regions expand until they
reach a  maximum size, after which  they collapse under  the action of
their own gravity.  It is within these haloes that gas cools and stars
are born.  This process may be studied using both analytical models or
numerical simulations,  in order  to understand just  what it  is that
determines    when    and    where    an    object    will    collapse
\citep{lacey93, lacey94, kauffwhite93, springel01b, giocoli07a}.

In the Spherical Collapse model (hereafter SC) \citep{gunn77},
which  describes the  evolution of  a spherical  mass shell within 
an expanding  background, the entire evolution is determined by the 
initial overdensity within the protohalo.  In this model, the initially 
spherical region remains spherical -- only its size changes -- and there 
is a critical initial overdensity $\delta_{sc}$ which a proto-halo must 
have for it to collapse and virialize by a given time (e.g. the present).  
Moreover, this value is the same for all haloes, whatever their mass.  

The  next  more  complicated  model assumes  an  Ellipsoidal  Collapse
(hereafter EC) (\citet{whitesilk1979}, \citet{bondmyers96}).  In the
formulation of  \citet{bondmyers96}, the triaxiality  in the potential
leads  to tidal  forces which  deform the  shape of  the object  as it
evolves.  Hence,  even if the  mass distribution of the  protohalo was
initially spherical, the final virialized object need not be a sphere.
Most implementations of the  EC model have assumed initially spherical
shapes  \citep{shethmotormen01,  shethtormen02, desjacques08, lamsheth08, 
rossi10},  though this is not  a fundamental assumption \citep{shen06}.
Since  the  EC  model  generically predicts  non-spherical  shapes  for
virialized objects, and simulated haloes are well-known to be triaxial
\citep{jingsuto02, allgood06},  the EC model is widely  expected to be
more realistic than the SC model.

Despite this, simulated haloes  are often still identified by searching
for  a critical  overdensity  within  a sphere,  or  deforming such  a
candidate spherical  region along its principal  axes.  Therefore, one
of the  goals of  this paper  is to present  a method  for identifying
ellipsoidal haloes which we  believe provides a more accurate estimate
of  halo masses  and  shapes.  We  then  explore how  our new  method
impacts estimates of the  resulting halo properties (e.g. mass, shape)
at the  present time,  and in the  initial conditions (e.g.,  mass and
deformation tensors of the  proto-halo).  We believe the latter allows
for more direct and quantitative tests of the EC model.

The outline of  this paper is as follows.  In Section~\ref{ECfacts} we
highlight those  aspects of the  Ellipsoidal Collapse model  which are
most relevant to our  study.  Section~\ref{sims} describes our dataset
and the  halo identification algorithm, explaining how  we improve our
estimates of  halo shapes, by  defining them as  virialized ellipsoids
rather than  spheres. In Section~4  we present our  results concerning
the difference in mass (Subsection  4.1) and in final shape (4.2) with
respect to the usual spherical one.  In Section~5 we study the initial
density  fluctuation  field  of   the  proto-haloes,  showing  how  the
difference  between  the  two  methods  is reflected  in  the  initial
properties  of  the  haloes  (Subsection  5.1); and  how  the  initial
properties influence  the subsequent evolution (5.2).  In Section~6 we
discuss the average evolution of halo shape and orientation.  A few 
case-by-case examples are presented in an Appendix.  We summarize and 
discuss our results and then conclude in Section~\ref{discuss}.

\section{The ellipsoidal collapse model}\label{ECfacts}
Many of the quantities we measure in the simulations are motivated by 
our desire to test various assumptions of the EC model.  So it is 
useful to summarize these here.  

\subsection{Does only potential matter?}
Although the EC model of \citet{bondmyers96} does not require it, 
all implementations of it assume that the mass distribution of a 
proto-halo is initially spherical.  In this case, the evolution of 
the shape is determined by just three numbers, which are specific 
combinations of the three eigenvalues of the initial $3\times 3$ 
deformation tensor centered on the center of mass of the proto-halo.  
(Strictly speaking the tensor is evaluated after smoothing the initial 
fluctuation field on the scale which contains all the mass which will 
end up in the halo.)  
Because the shape of the object is determined by the initial deformation 
tensor, in this model, the mass and deformation tensors are perfectly 
aligned {\em by definition}.  Measurements in simulations have shown 
that this vast simplification is actually a rather good approximation 
\citep{porciani2002b}.  This means that tidal torques, induced by the 
misalignment between these tensors, are subdominant.  

Therefore, in this model, the proto-halo shrinks (in comoving coordinates) 
fastest along the axis corresponding to the direction of maximum 
compression, forming a pancake; this is followed by collapse along 
the second and then finally along the axis of least compression 
eventually resulting in a virialized object.  I.e., the three axes 
collapse at different times, which depend on the local shape of the 
deformation tensor, but because the initial shape of the mass 
tensor was spherical, the ordering of the axis lengths is monotonic 
in time, and collapse times are determined by the initial compression 
factors:  larger compression factor means earlier collapse.  

A final assumption of the model is that virialization corresponds to 
collapse along all three axes.  Since the first axis to collapse 
and freeze-out from the expansion of the background universe 
will have done so when the universe was denser, this first axis to 
collapse will also be the shortest axis of the virialized object, 
and the third will be longest.  
I.e., the shortest axis at virialization will be aligned with the 
direction of initial maximum compression, and vice versa.  

The three numbers which determine the evolution in this model are 
the initial density  contrast $\delta_{i}$ (the only number that 
matters for the spherical evolution model), and the ellipticity and 
prolateness parameters $e$ and $p$ of the deformation tensor (i.e. 
not the mass tensor).  These are defined as follows.  If the 
eigenvalues of the deformation tensor are
 $\lambda_1\ge \lambda_2\ge \lambda_3$, 
then 
\begin{equation}
 \delta_{i} \simeq \lambda_{1}(t_i) + \lambda_{2}(t_i) + \lambda_{3}(t_i)
\end{equation}
and 
\begin{equation}
  e = \frac{\lambda_{1}(t_{i})-\lambda_{3}(t_{i})}{2\delta(t_{i})}
  \qquad \mathrm{and} \qquad
  p = \frac{\lambda_{1}(t_{i})+\lambda_{3}(t_{i})-2 \lambda_{2}(t_{i})}{2\delta(t_{i})}
\end{equation}
where $e\geq 0$  and $|p|\leq e$; 
a sphere has $e=p=0$, prolate configurations have $-e\leq p \leq 0$, 
and $0\leq p \leq e$ are oblate.  

\citet{shethmotormen01} showed that, in this model 
\begin{equation}
  \frac{\delta_{ec}(e,p)}{\delta_{sc}} = 
  1 + \beta\left[5(e^{2}\pm p^{2})\frac{\delta^{2}_{ec}(e,p)}{\delta^{2}_{sc}}\right]^{\gamma},
 \label{decep}
\end{equation}
where  $\beta= 0.47$  and  $\gamma= 0.615$.   Note that  non-spherical
effects always make $\delta_{ec}(e,p) > \delta_{sc}$, a point to which
we  will  return.   \citet{shethmotormen01}  also showed  that,  in  a
Gaussian  random fluctuation  field,  $\delta,e,p$ can  vary from  one
position to  another, with  the consequence that  even at  fixed mass,
$\delta_{ec}$ can vary from one proto-halo to another.  They then used
the  statistics  of  Gaussian   fields  to  argue  that,  on  average,
$\delta_{ec}$ will be  close to $\delta_{sc}$ at large  masses, but it
will  increase  as  mass   decreases.   They  showed  that  this  mass
dependence of  $\delta_{ec}$ was indeed evident  in their simulations,
and so it is now commonly  assumed that the EC model describes some of
the physics  which is  relevant to the  triaxial shapes  of virialized
haloes in simulations.  In what follows, we test this in slightly more
detail by checking if the dependence on $e$ and $p$ is as predicted.

\subsection{What if the initial shape is not spherical?}
However, we  already know that  there is one  respect in which  the EC
model  fails.  This  is  because  the model  predicts  that the  large
proto-halo patches which collapse to form massive haloes should be more
spherical.  Therefore, the shape  at virialization should also be more
spherical \citep{rossi10}.   Simulations show that, in  fact, the most
massive     virialized     haloes     can     be     quite     triaxial
\citep{jingsuto02,allgood06}.  This raises the question of whether the
EC  model has failed  to identify  the correct  shapes in  the initial
conditions, or if its approximation of the evolution is incorrect.

For  example, the next  simplest model  would begin  with proto-haloes
with triaxial rather than  spherical mass distributions, but will keep
the  assumption that the  mass and  deformation tensors  are perfectly
aligned.  In this case, the evolution of the shape, and hence the time
required to collapse and virialize  depends on $\delta,e,p$ as well as
on the initial axis lengths (i.e., the square-roots of the eigenvalues
of the  mass tensor), and on  whether or not the  direction of maximum
compression is  oriented along the  initially longest or  the shortest
axis.  In  such a model, one  might postulate for  example that haloes
form at those special points in  space where all three axes would have
turned around or collapsed at the  same time; this would happen if the
axis which was initially the  longest also had the largest compression
factor initially (`perfect'  alignment).  This correlation between the
directions  of the longest  initial axis  and the  largest compression
factor is indeed  seen in simulations \citep{porciani2002b}, in the 
sense that $\cos\theta_{11}\sim 1$ where $\theta_{11}$ is the angle 
between these two vectors.  

In  fact,   despite  the  theoretical   simplicity  of  these `perfect'
alignment models  , the evolution of  the shape can appear  to be more
complex because, e.g., if the axis that was longest initially also had
the largest compression factor,  then it may eventually become shorter
than what  was initially the second  longest axis.  So  one might ask,
even  though the  shortest  axis  initially may  not  be the  shortest
finally,  is it  still true  that the  shortest axis  at virialization
tends to be aligned with  the direction of initial maximum compression
(and vice versa)?   If it does, then this would  still be in agreement
with an  EC model for the  evolution, only applied  to a non-spherical
initial mass distribution.

The Zel'dovich approximation \citep{zeldovich70}, provides an easy way
to see  the qualitative features  discussed above.  In this  case, the
eigenvalues    are   assumed    to   evolve    as    $\lambda_j(t)   =
\lambda_j(t_i)\,D(t)/D(t_i)$, where $D(t)$ is the linear theory growth
factor at time $t$.  This  means that the comoving axis lengths evolve
independently   of   one   another,   as  $R_j(t)   =   R_j(t_i)[1   -
\lambda_j(t)]$.   Notice   that  in  this   approximation  a  positive
eigenvalue implies contraction,  whereas a negative eigenvalue implies
expansion.

The nonlinear density is the ratio of the mass of the proto-halo to 
its volume.  Mass conservation means that it satisfies 
\begin{equation}
  1+\delta(t) = \frac{M}{\bar\rho V(t)}
              = \frac{1}{\prod_{j=1}^3 1-\lambda_{j}(t)}.
\end{equation}
Notice that this expression applies even if the $R_j(t_i)$ are not 
equal; it reduces to $\delta_{i}=\sum_{j=1}^3 \lambda_{j}$ 
for very small values of $\lambda_{i}$.  

In this approximation the collapse of axis $j$ corresponds to 
the time when $1-\lambda_j(t)=0$.  When the $R_j(t_i)$ are all equal, 
then the order of the axis lengths $R_j$ at any given time is 
determined completely by the ordering of the $\lambda_j$ at the 
initial time.  But if they are not equal, then, although the order 
of the time to complete collapse is still determined by the order 
of the $\lambda_j$, the axis lengths at some time prior to complete 
collapse may not.  E.g., if $R_j > R_k$ {\em and} $\lambda_j>\lambda_k$ 
then $R_j$ must collapse before $R_k$, so at some point it must become 
smaller than $R_k$.  
Notice that, despite this reordering of the axis lengths, the 
three fundamental directions of the principal axes of the mass 
tensor will {\em not} have changed.  Of course, this means that 
if one studies the direction in which `the longest axis' points, 
then this direction may change suddenly (e.g., at the time when it 
stops being the longest).

Although \cite{shen06} provide a simple approximation to the 
evolution predicted by the EC model when the initial shape is not 
a sphere, which can be thought of as a simple physically motivated 
modification to the Zeldovich approximation, we believe that a 
complete understanding of these and related aspects of the EC model 
is still missing.  For example, we noted that $\cos\theta_{11}\sim 1$, 
where $\theta_{11}$ is the angle between the longest axis and the 
direction of maximum compression.  While we confirm that 
$\cos\theta_{11}\sim 1$, the torque generated by the misalignment 
between these two vectors depends on $\sin\theta_{11}$, and it can 
be quite different from zero even though $\cos\theta\sim 1$ (basically 
because $\cos 30^\circ \sim 0.866$ but $\sin 30^\circ\sim 0.5$).  
Thus, even a small misalignment can have a dramatic effect, so 
further study of slightly misaligned models is certainly needed.  

Even so, a comparison between simplified models such as these and 
simulations clearly depends critically on correctly identifying 
the proto-haloes in the initial conditions, and this in turn 
depends on identifying the particles which belong to the final 
virialized haloes.  Therefore, in what follows, we analyse in detail 
the shapes of haloes, how these depend on the method of estimation, 
how they evolve, and how this evolution is influenced by the potential 
field.  

\section{Simulation data}\label{sims}
We address these questions by studying the properties of haloes 
identified in the GIF2 simulation.  

\subsection{The GIF2 simulation}
The GIF2  Simulation \citep{gao04} adopts  a $\Lambda$CDM cosmological
model with  $\Omega_{m}=0.3$, $\Omega_{\Lambda}=0.7$, $\sigma_{8}=0.9$
and $h=0.7$.   It follows  $400^{3}$ particles in  a periodic  cube of
side $110h^{-1}$  Mpc from an  initial redshift $z=49$ to  the present
time.  The  associated change  in the linear  theory growth  factor is
$D_{+}(z=0)/D_{+}(z=49)=38.993$.   The  individual  particle  mass  is
$1.73\times  10^{9}h^{-1}$   $M_\odot  $.   Initial   conditions  were
produced  by  imposing perturbations  on  an  initially uniform  state
represented  by a glass  distribution of  particles \citep{white1996};
based on the  Zel'dovich approximation \citep{zeldovich70}, a Gaussian
random field  is set up by  perturbing the positions  of the particles
and assigning them velocities  according to the growing model solution
of  linear  theory \citep{cmbfast1996}.   The  critical  value of  the
linear theory  overdensity that is required for  spherical collapse at
the present time is $\delta_{c}=1.6755$.

\subsection{Halo identification}
Dark  matter haloes  are identified  as  local maxima  in the  density
field: this  is commonly done  by growing spherical shells  around the
halo center  - defined for example  by the position of its most bound
particles - and finding some  characteristic radius $R$ where the mass
overdensity reaches a given value: $\delta_{vir} \equiv (\rho(<R) - \rho_b)
/  \rho_b$,  with  $\rho_b$  the  mean density  of  the  universe  and
$\rho(<R)$ the  average density inside radius $R$.  Common choices for
this {\em virial} overdensity  are $200$ times the background density,
$200$  times the  critical density,  and the  value prescribed  by the
spherical  collapse  model:  $\delta_{vir}=323.7$  (at $z=0$  for  the
assumed cosmological  model - \citet{eke1996}); this algorithm
is known as Spherical Overdensity (SO).

Another  method  for identifying  collapsed  structures  uses a
percolation  algorithm,  called Friends  of  Friends  (FOF), that  links
together  particles  closer   than  $b  d$,  where  $d$  is  the  mean
interparticle separation and $b <1$  is called {\em linking length}; 
$b=0.2$ is a common choice, as it identifies structures with average  
density of  order one  hundred times ($\sim 1/5^3\times$) the  background.  
$FOF$ haloes are  usually not spherical,  but closely follow  the isodensity
contours  of  the  identified  structures,  giving  a  more  realistic
representation of the  halo shape.  On the other  hand, this algorithm
may link together haloes  which are dynamically different, if particle
noise is such  that there exists a thin  bridge of particles connecting
them.

A third and far less commonly used option is to define  haloes as triaxial
ellipsoids, as we do in the present work.  Specifically, we will define
haloes  as  triaxial  structures   with  mean  overdensity  $\delta  =
\delta_{vir}$.   Although  computationally  more costly,  this  method
tries to retain the advantages  of both previous ones: a theoretically
motivated virial overdensity value and a more realistic description of
the actual halo shape. This description is more consistent
with  the EC  model, which naturally predicts triaxial rather than 
spherical haloes. In  the next  Subsection we  will  describe our
algorithm in detail.

\subsection{Triaxial haloes}
The volume $V$ of a  triaxial ellipsoid is defined by $V=(4\pi/3)abc$,
with  $a, b$,  and $c$  the longest,  intermediate and  shortest axis,
respectively.  In order to find  the ellipsoidal shape which best fits
a given  halo, we first run  an $SO$ algorithm on  the full simulation
and find for  each halo its virial radius  $R_V$, enclosing an average
overdensity  $\delta_{vir}$   \citep{tormen04,  giocoli08b}.  We  then
calculate  the  mass tensor  $M_{\alpha\beta}$  defined  by the  $N_V$
particles found inside $R_V$ as:
\begin{equation}
 M_{\alpha\beta} = \frac{1}{N_V}\sum_{i=1}^{N_{V}} r_{i,\alpha}r_{i,\beta}
\end{equation}
where $\textbf{r}_{i}$  is the position  vector of the  $i$th particle
and $\alpha$ and  $\beta$ are the tensor indices.   Note that, even if
the halo  distribution can be  recovered from both, this  is different
from the inertia tensor which is defined as:
\begin{equation}
I_{\alpha\beta}=\sum_{i=1}^{N} m_{i}(\textbf{r}^{2}_{i}\delta_{\alpha\beta}-r_{i,\alpha}r_{i,\beta}).
\end{equation}
As explained  in \citet{bett07}, much  of the literature  confuses the
two  tensors  and  uses  both  interchangeably to  describe  the  mass
distribution.

The mass tensor  so found will not be isotropic even for particles 
within a sphere if the particle distribution inside $R_V$ is not
isotropic. Therefore, by diagonalizing $M_{\alpha\beta}$ we will obtain 
eigenvalues and eigenvectors which give  an initial guess for the true
shape and  orientation of  the virialized structure:  the axes  of the
best fitting  ellipsoid are  defined as the  square roots of  the mass
tensor eigenvalues.  We then modify the list of particles which make 
up the halo by performing  a sort of  \emph{Ellipsoidal Overdensity} 
criterion: for each  particle selected in  the previous  step, we 
calculate its ellipsoidal distance from the center as
\begin{equation}
 r^{3}_{E} = \frac{\Delta x^{2}}{l^{2}_{1}}
          + \frac{\Delta y^{2}}{l^{2}_{2}} + \frac{\Delta z^{2}}{l^{2}_{3}},
\end{equation}
where $l_{1}^{2}$, $l_{2}^{2}$ and  $l_{3}^{2}$ are the eigenvalues of
the mass tensor calculated at  the previous step.  (The eigenvalues of
the  inertia tensor  would be  $l_2^2 +  l_3^2$, $l_1^2  +  l_3^2$ and
$l_1^2 + l_2^2$.)  Sorting  the particles by ellipsoidal distance from
the center  of the halo,  we build up  an ellipsoid which  encloses an
average  overdensity  $\delta_{vir}$.    We  believe  that  using  the
overdensity  to  select  the   halo  particles  is  more  precise  and
consistent than  requiring the volume  of the ellipsoid equal  that of
the original  SO sphere, or requiring  the longest axis  equal that of
the  initial sphere  \citep{warren92,  allgood06, schneider12}.   This
also  allows a  more direct  comparison with  theoretical  models.  We
recalculate the  mass tensor for  this new particle  distribution, and
obtain  a  new  set   of  eigenvectors,  which  improve  the  previous
description of the halo shape.   We iterate this calculation until the
algorithm  converges to  a  set  of eigenvectors  to  better than  one
percent in the axial ratios. We  choose only the haloes with more than
200  particles to  ensure a  good resolution  and we  removed  all the
unbound haloes (which  were about 6\% in the  original sample of 15001
haloes);  for  some of  the  plots  we  use other  selection  criteria
(described throughout the text),  to refine our subsample depending on
the specific  needs of the analysis.   The final result is  a new halo
catalogue which is the one we study in the remainder of this work.

\begin{figure}
\includegraphics[width=9cm]{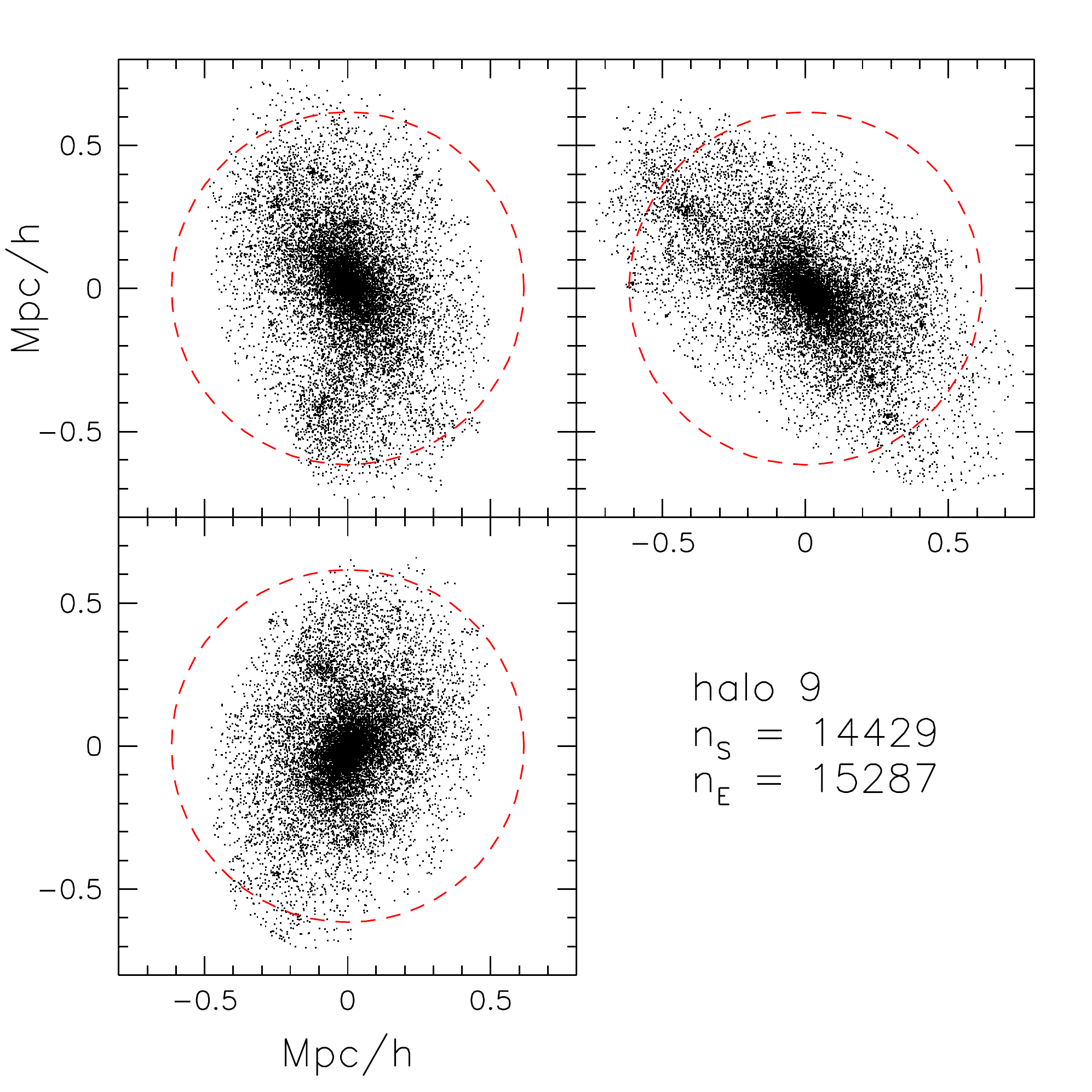}
\caption{Example of the difference  between the final ($z=0$) spherical
  or ellipsoidal identification for a halo of the simulation: we show the
  projected  distribution (in  the \emph{x-y},  \emph{y-z}, \emph{x-z}
  planes) of the halo particles  inside the ellipsoid; the red dashed
  circle   indicates   the  radius   of   the   halo   in  the   spherical
  identification. We note that the shape becomes in general more
  elongated and that the number of particles increases, indicating
  that the ellipsoidal shape traces the isodensity contours better
  than the spherical one. \label{haloexample}}
\end{figure}

\section{Haloes at $z=0$}

Figure  \ref{haloexample} illustrates the  difference between a 
spherical halo and its triaxial counterpart: black dots show the  
projected  distributions  of  halo  particles  inside  the  final
ellipsoid,  while the red  dashed circle  indicates the  virial radius
$R_V$.  The triaxial  shape is more elongated when  measured inside an
ellipsoidal volume (in this case particularly in the \emph{y-z} plane)
and follows  the natural  orientation of the  halo and  its isodensity
contours more accurately than does the  $SO$ halo. We expect, therefore, 
an  increase in virial mass  $M$: this is true  for the test halo of 
Figure \ref{haloexample} and  in the next Section we will show results 
for the whole halo sample.

\begin{figure}
\includegraphics[width=8cm]{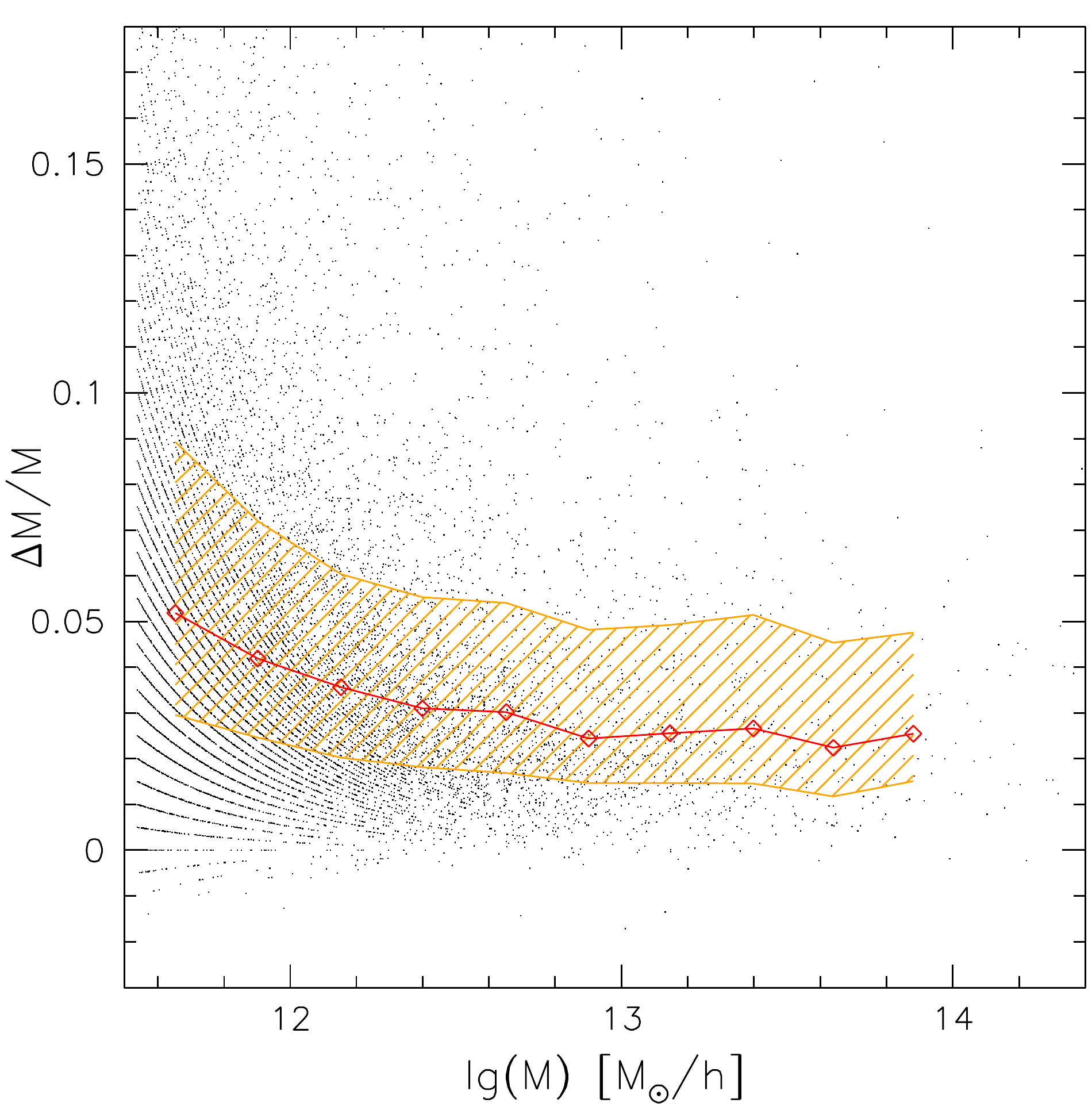}
\caption{The mass difference  between  the ellipsoidal  and
  spherical  identification methods, shown as  a  function  of  the
  spherical mass.  The medians of the distribution are shown in red and
  the region which lies between the first  and third quartiles is 
  shaded orange. \label{masschange}}
\end{figure}

\subsection{Difference in mass}

In Figure  \ref{masschange} we plot  the fractional  difference  in 
mass  between  the ellipsoidal  and spherical identifications, 
defined by the ratio
\begin{equation}
 \frac{\Delta M}{M}=\frac{M_{E}-M_{S}}{M_{S}}
\end{equation}
as a function of the spherical mass $M_S$. Notice that our new method 
indeed gives  a different  estimate of  the mass:  at low  masses 
the median difference of about 5\% is larger  than at high masses.  
This indicates that small mass  haloes are more elongated  and  so  
the SO estimate is more biased than it is at larger masses.  
The ellipsoidal  mass is clearly larger than the spherical  one, 
as ellipsoids indeed trace the isodensity surfaces more precisely, 
and so include more particles.  Of course the  difference between 
the two estimates  cannot be  too big, since the ellipsoidal mass 
is  a refinement of the spherical one.  This has a positive 
implication: the spherical overdensity criterion, which is simpler 
than our best  fitting ellipsoid method and requires fewer 
calculations, turns  out to  be quite precise  in estimating  the halo
mass and thus the halo mass function.  However, as we will show in the
next section,  the ellipsoidal description is much more appropriate 
for estimating halo shapes. There is a small fraction of 
haloes (less than 1\%) which have negative  values of $\Delta M/M$:
by individual inspection, we found that they all correspond to merging
haloes.  Moreover,  the initial (Lagrangian space) distribution of 
their  particles is often fragmented  in two or more large  regions. 
We exclude them from further analysis.  

\subsection{Difference in final shape}
\begin{figure}
\includegraphics[width=8cm]{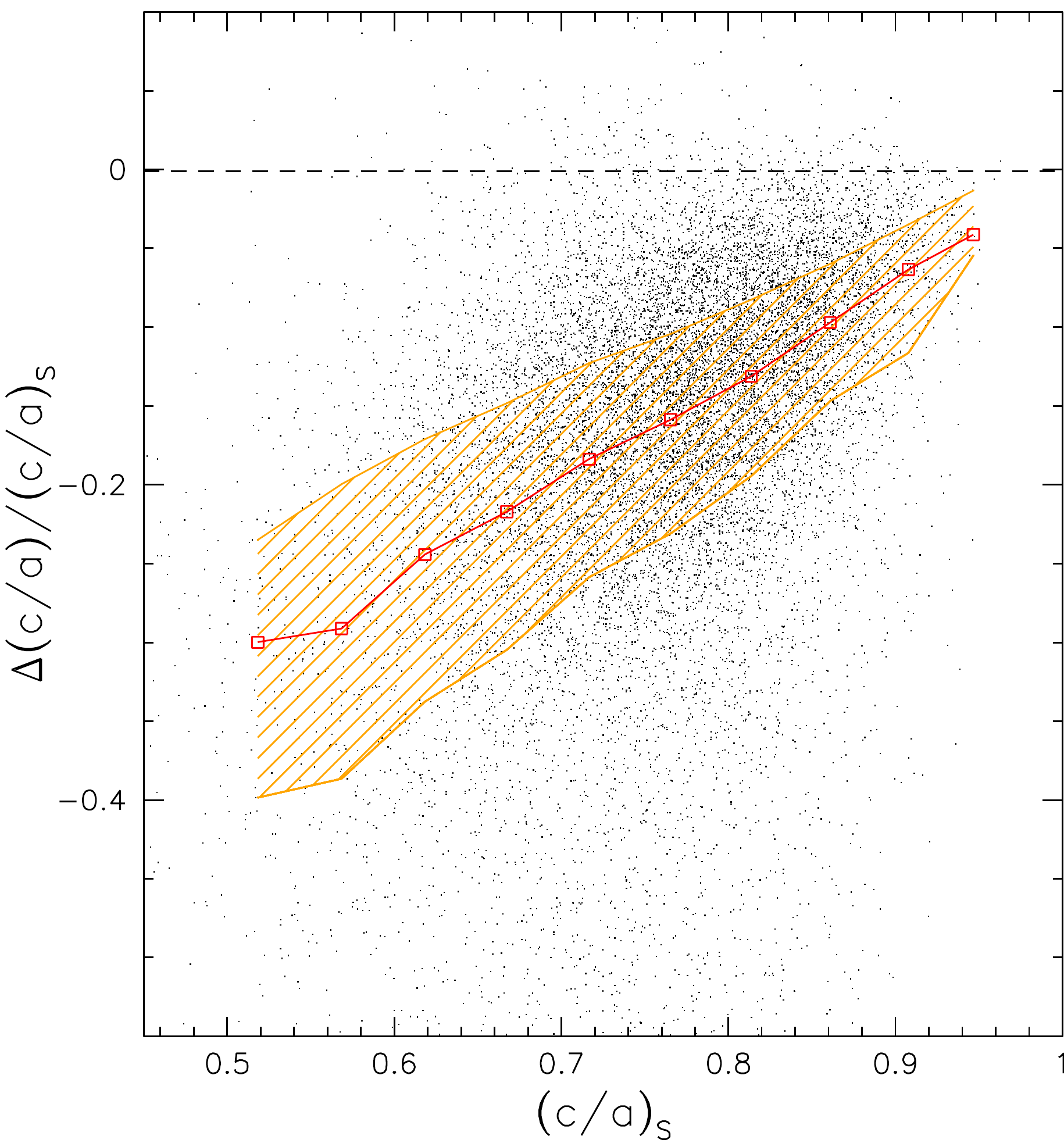}
\includegraphics[width=8cm]{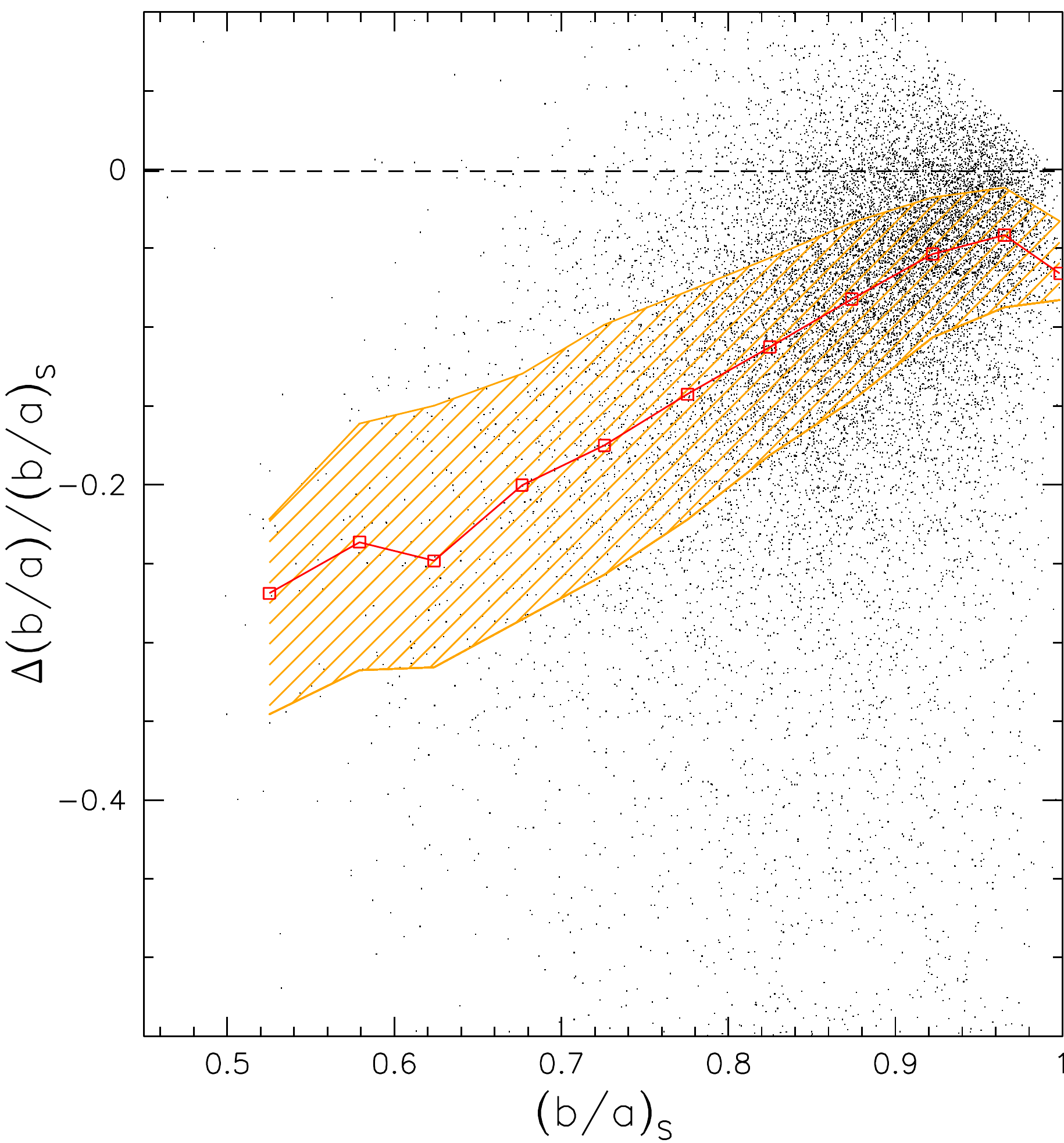}
\caption{The difference  in the final  axial ratios  \emph{c/a} and
  \emph{b/a} ($a\leq b\leq  c$), as a function of  the spherical
  ones. The orange  shaded region lies between the first and third
  quartiles; red points show the  median.  The  relative
  difference is  generally negative, indicating that fitting 
  ellipsoids yields more elliptical and prolate shapes than 
  fitting spheres. 
  \label{axeschange}}
\end{figure}

\begin{figure}
\includegraphics[width=8cm]{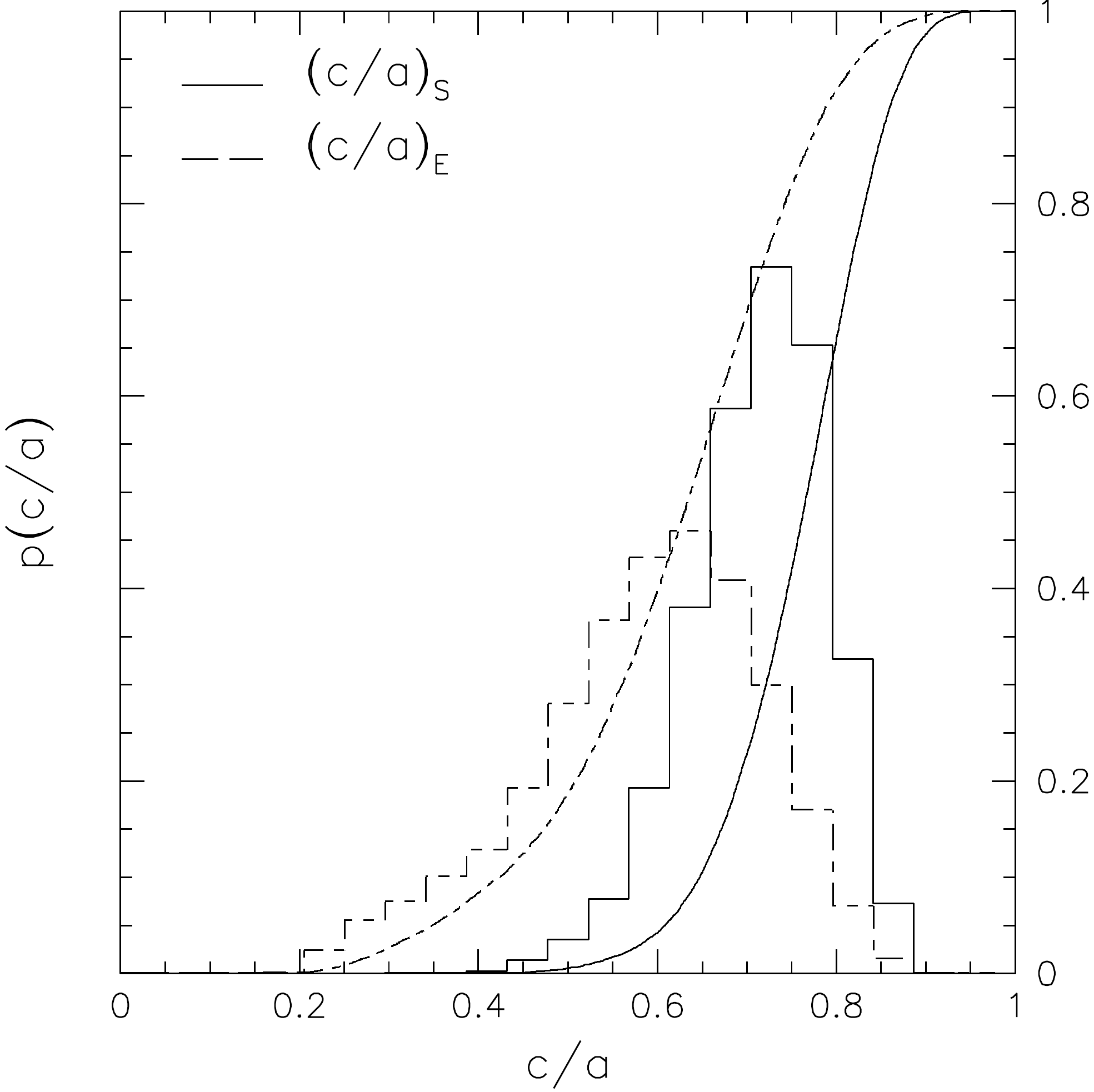}
\includegraphics[width=8cm]{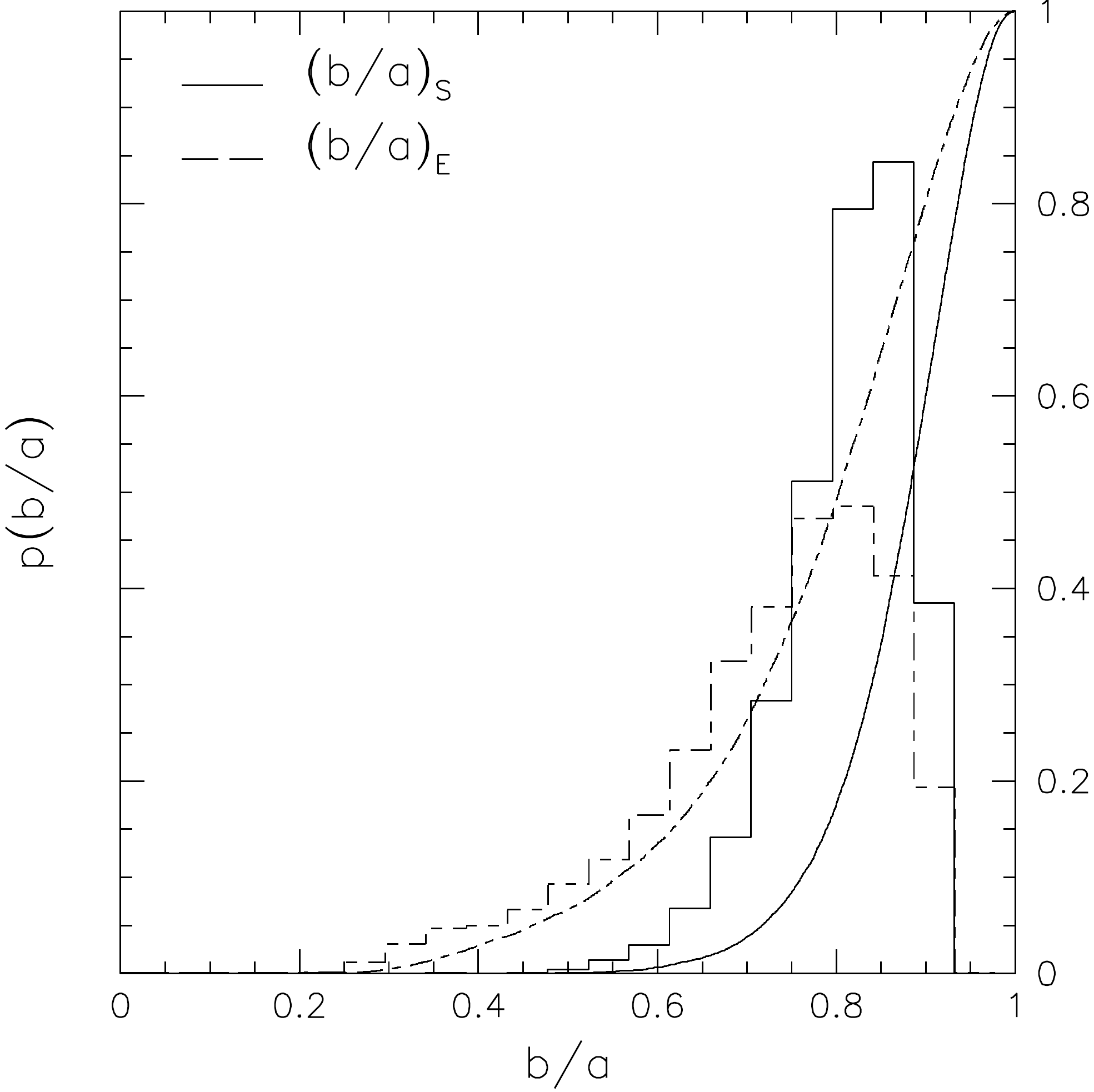}
\caption{Cumulative  and  differential  distribution of  axial  ratios
  \emph{c/a} and  \emph{b/a} at the shape corresponding  to the virial
  overdensity,  from corresponding to  a fitting  ellipsoidal (dashed)
  and  spherical (solid)  volume: in both cases the enclosed mass
  distribution is not isotropic and so we are able to calculate the
  axial ratios.   Haloes identified  with the  $EO$
  criterion  have  a smaller  median  value,  with significantly  more
  objects  at  small $b/a$  and/or  $c/a$.   This  difference is  more
  significant for \emph{c/a}, meaning  that haloes become more prolate
  when fit with ellipsoids.  \label{axesdist}}
\end{figure}

We  now   study  halo  shapes.   Figure   \ref{axeschange}  shows  the
fractional difference in the axial ratios $c/a$ and $b/a$ (with $a\geq
b\geq c$ the  three eigenvalues of the mass tensor),  as a function of
the spherical axial ratios using the fractional differences
\begin{equation}
 \frac{\Delta(c/a)}{(c/a)_{S}} = \frac{(c/a)_{E}-(c/a)_{S}}{(c/a)_{S}} 
\end{equation}
and  similarly for  $b/a$.  Ellipsoidal  haloes are,  of  course, more
ellipsoidal, so  $\Delta(c/a)/(c/a), \Delta(b/a)/(b/a)\leq 1$  .  When
the axial ratio  approaches $1$, then haloes are  almost spherical, so
the  ellipsoidal  method correctly  returns  a  spherical shape.   The
difference  increases  as  $(c/a)_{S}$  and $(b/a)_{S}$  decrease:  in
particular,  ellipsoidal-based shapes become  more prolate,  since the
shortest  axis  changes  more than  the  others,  as  can be  seen  by
comparing the  two panels of Figure \ref{axesdist}.   This agrees with
the recognised preference for prolate halo shapes in previous works.

In Figure   \ref{axesdist}  we  show   the  differential   and  cumulative
distributions  for both  $c/a$  and $b/a$:  the distribution  obtained
using final spheres is represented by the solid line, while the dashed
line is for ellipsoids.  There are more objects with small $c/a$ when 
fitting ellipsoids, and the median is shifted to smaller values.  
This is also true for the   $b/a$  distribution, although   the  
differences are  smaller, implying a  preference for more prolate shapes. 
This confirms what we already noticed in Figure~\ref{axeschange}.
A similar  result was obtained  by \citet{warren92}, who  compared the
axial ratios of a spherical and an ellipsoidal halo, finding a change
in the axial ratios in the direction of a more ellipsoidal shape.

It is also  interesting to study the halo structure  at a given time,
to check if (a) the shape depends on mass and (b) if, at fixed mass, 
the shape varies as a function of distance from halo center.  
We find that massive haloes are less round than lower mass haloes 
since this agrees with previous work \citep{jingsuto02, allgood06}, 
we do not show plots here.  
Our analysis also confirms the general opinion that, on  average, 
there is also a systematic dependence of shape on radius:  
haloes are more elongated closer to the center, and more spherical in 
the outskirts \citep{allgood06, jingsuto02, warren92}. 
In addition, more massive haloes have a steeper gradient in axial 
ratios than lower mass haloes:  this happens  because the more massive 
haloes  are less  influenced by  their environment.

We also measured the misalignment between the axis direction at a 
certain fraction of the radius and the axis at $0.5r_{vir}$.  
For this, we choose a subsample of the haloes, restricting our 
analysis to those for which the axes direction is determined more
precisely. For this purpose we define the accuracy in the axis 
determination by 
\begin{equation}
 \theta_{err} = \frac{1}{2\sqrt{N}}\frac{\sqrt{r}}{1-r}\,{\rm rad}
             < 0.1\, {\rm rad}
\end{equation}
where $N$  is the  number of  particles used and  $r$ is  the relevant
axial  ratio: $b/a$  for the  major axis,  $a/b$ for  the minor  axis, and
max[$b/a$; $a/b$] for the intermediate  axis \citep{bailin04, bailin05}. 
Notice that the alignment is slightly better for more massive haloes, 
represented  by the black curve,  but on  average  the internal  
alignment  is very  good, confirming  previous  results  
\citep{jingsuto02, bailin05, Vera-CiroAq11, schneider12}.

\section{Protohaloes and the initial field} 
We have  discussed the  differences between spherical  and ellipsoidal
haloes, comparing  the halo properties  at the present time,  when the
haloes are  collapsed and virialized.  We now study these differences 
for proto-haloes in the initial conditions.  

The protohalo  regions are  defined  by  tracing   all  halo  particles, 
identified at $z=0$, back to their unperturbed (Lagrangian) positions.
For  each  protohalo we  calculate  the  elements  of the  deformation
tensor, defined at each  position \textbf{q} as the second derivatives
of  the gravitational potential  $\Phi$.  This,  in the  the Zel'dovich
approximation \citep{zeldovich70}  is equivalent to  the evaluation of
the first derivatives of the initial displacement:
\begin{equation}
\xi_{ij}(\textbf{q})= -\frac{\partial^{2}\Phi}{\partial x_{i}\partial
  x_{j}}(\textbf{q})=-\frac{\partial\Psi}{\partial x}.
\end{equation}
These  were   calculated  from  the  initial   displacement  grid  and
differentiated with  respect to the spatial coordinates.  

Specifically, for each halo we flagged the grid points occupied by
particles and calculated the deformation tensor as:
\begin{equation}
 \xi_{ij} = \frac{1}{V_{L}}\int_{V_{L}}\xi_{ij}(\textbf{q}) d^{3}x 
         = \frac{1}{N_{G}}\sum_{k=1}^{N_{G}}\xi_{ij}(k)
\end{equation}
where $N_{G}$  is the sum of  all the grid cells  contained within the
lagrangian  volume  of  the  halo: i.e., those actually  occupied  by  halo
particles  and those  left empty,  but still  located inside  the halo
(with at least four neighbor  cells occupied by particles). These last
must be considered since the  total potential field acting on the halo
is affected by their  contribution. Thus, we used an algorithm to
select  the correct set of  empty cells  and  added their  contribution 
to  the deformation tensor of the halo.  This results in a small change 
to the original value, which refines the one obtained using the particle 
grid points only. Since the shape of the protohalo regions is not 
symmetric nor regular, we could not choose a characteristic radius and
use it to smooth the distribution in Fourier space (and calculate the
value od the deformation tensor from it), since we need to maintain
spatial resolution at each point.

While  doing this,  we  also studied  the  profiles of  haloes in  the
initial conditions.   Since the particles  belonging to each  halo are
selected at  the present time, they  do not necessarily  form a single
simply connected \emph{lump} in the initial conditions.  We found that
haloes are  indeed more fragmented  in the initial conditions,  but in
most cases the  mass of the main  lump is more than 90\%  of the total
mass  of the  halo.  Thus,  to  calculate the  initial properties,  we
decided  to use  the  particles contained  in  the main  lump, and  to
exclude from our  sample those haloes for which less  than 90\% of the
mass is in  the main lump.  (In some rare cases,  this fraction can be
as small as  50\% of the mass: these haloes  probably formed through a
recent merging process or have undergone some transformations in time,
so  they do not  correspond to  a sufficiently  compact region  in the
initial conditions.  They  are only 8\% of the  total sample: removing
them does not affect the mean behavior, but it does reduce the scatter
around this  mean.) Summing up,  $N_{G}$ is the  sum of all  the cells
(full and empty) contained in the main lump of the each halo.

\begin{table}
\centering
\begin{tabular}{|c|c|c|}
  \hline
  $\nu$ & mass & n \\
  \hline
  0.53 & $M_{\star}/16$ & 6607 \\
  0.67  &
  $M_{\star}/4$ & 3890 \\
  0.84 &
  $M_{\star}$ & 1061 \\
  1.06 &
  $2-4M_{\star}$ & 348 \\
  1.33 &
   $8-16M_{\star}$ & 109 \\
  \hline
\end{tabular}

\caption{Correspondence between mass bins and $\nu$; 
  $M_{*}$ is $8.956\times 10^{12}M_{\odot}h^{-1}$.
          The third column shows the number of haloes in each bin.
          \label{tab_bins}}
\end{table}

\begin{figure}
\includegraphics[width=9cm]{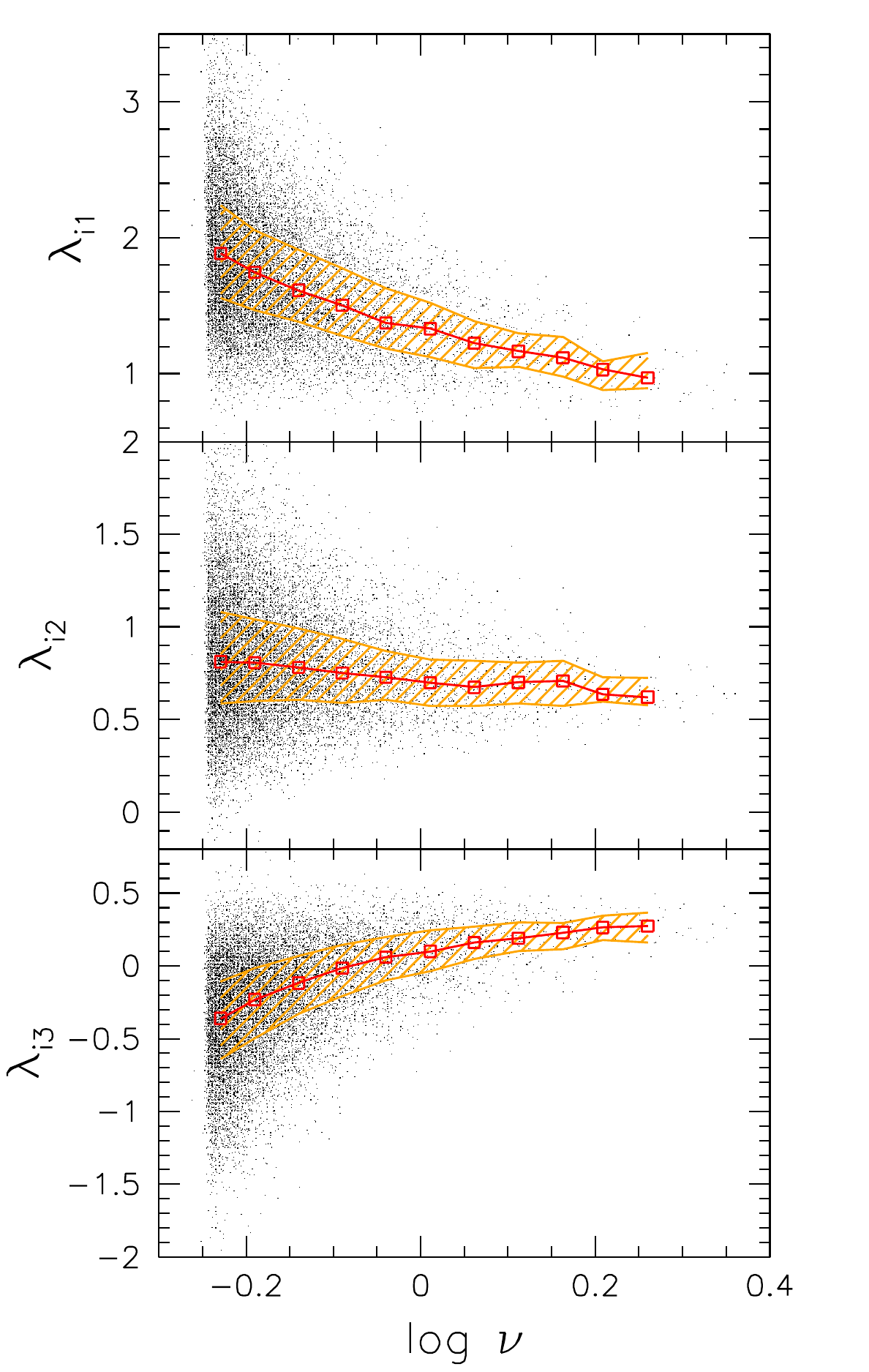}
\caption{The eigenvalues of the deformation tensor for the ellipsoidal
  proto-haloes,  as a  function  of  mass (which  has  been scaled  to
  $\nu$). The median  of each distribution is shown  in red, while the
  orange shaded region lies between the first and the third quartiles.
  Note that, while  the first two eigenvalues are  positive, the third
  one is  almost always negative,  especially at low  mass, indicating
  that haloes are not contracting with the same strength along all the
  three directions and that the potential along the third axis may
  slow down the contraction in that direction.\label{aut_def}}
\end{figure}

\subsection{Non-positivity of the initial eigenvalues}
Figure~\ref{aut_def} shows the distribution  of the $\lambda_{i}$ as a
function  of  halo mass,  which  we express  in  scaled  units $\nu  =
\delta_{sc}/\sigma(M)$, where $\sigma^2(M)$ represents the variance in
the  initial density  fluctuation field  when smoothed  on scale  $R =
(3M/4\pi\bar\rho)^{1/3}$ (the $\lambda_{i}$s  are rescaled to $z=0$ to
allow an easier comparison with the distribution of $\delta_{i}$ shown
in  Figure \ref{deltaec}).   In principle,  this expression  assumes a
spherically symmetric filter.  Although \citep{lamsheth08} provide the
relevant expression  for non-spherical filters, and note  that it only
makes a small difference, our main purpose here is to rescale the mass
variable, so  using the spherical expression is  appropriate. In these
units,  $\nu=1$   corresponds  to  a  mass   $M_{*}$  is  $8.956\times
10^{12}M_{\odot}h^{-1}$.  Later on, we  will study halo properties for
broader bins  in mass;  Table \ref{tab_bins} shows  the correspondence
between these mass bins and $\nu$.

Figure~\ref{aut_def}  shows  that  while  the  first  and  the  second
eigenvalues are  mostly positive,  the third one  is negative  for the
majority of  the haloes, especially  at lower masses.   This indicates
that,  on  average, protohaloes  in  the  initial  conditions are  not
contracting with the same strength  along all their axes, and that, in
the direction of $\lambda_{3}$ the  potential may act slowing down the
contraction.   (We   note  in  passing   that  $\lambda_3<0$  directly
contradicts the fundamental assumption of \citet{leeshan98}'s model of
halo abundances,  meaning their model is  untenable.)  More precisely,
only   $30\%$  show   $\lambda_{1,2,3}\geq  0$,   while   $70\%$  have
$\lambda_{3}\leq 0$  and $\lambda_{1,2}\geq 0$.  A  small fraction ($<
1\%$) of haloes behave very differently from these two categories.

So  far  we  have  checked  the differences  between  ellipsoidal  and
spherical identification as far as mass and final shape are concerned.
It is natural to ask how the corresponding initial deformation tensors
differ.   We found  that, as  for  the  mass, the  two identification 
schemes yield very  similar results: although there are haloes  
(especially at  low mass)  with significant  differences, even
larger than 40\%, the median of the distributions remains around zero.
The variance  is larger in the  case of $\lambda_{3}$, and, of course 
for lower masses.   This shows that the spherical  overdensity criterion 
traces the potential (and  so the mass) quite well and that the
best fitting  ellipsoid is  useful mainly for  the description  of the
geometrical  shape  of the  haloes  and  its  evolution.  The  greater
discrepancy for the values of  $\lambda_{3}$, which is the one showing
an  unexpected behavior, suggests  that our  new procedure  provides a
more realistic description of the halo structure.

\subsection{Protohalo overdensities:  The trace of the initial deformation tensor}
In the EC model, the fundamental quantities which influence the 
evolution are not the eigenvalues of the deformation tensor 
themselves, but the combinations $\delta,e,p$.   The most important 
of these is the overdensity $\delta$, which is the trace of the 
deformation tensor.

\begin{figure}
\includegraphics[width=7cm]{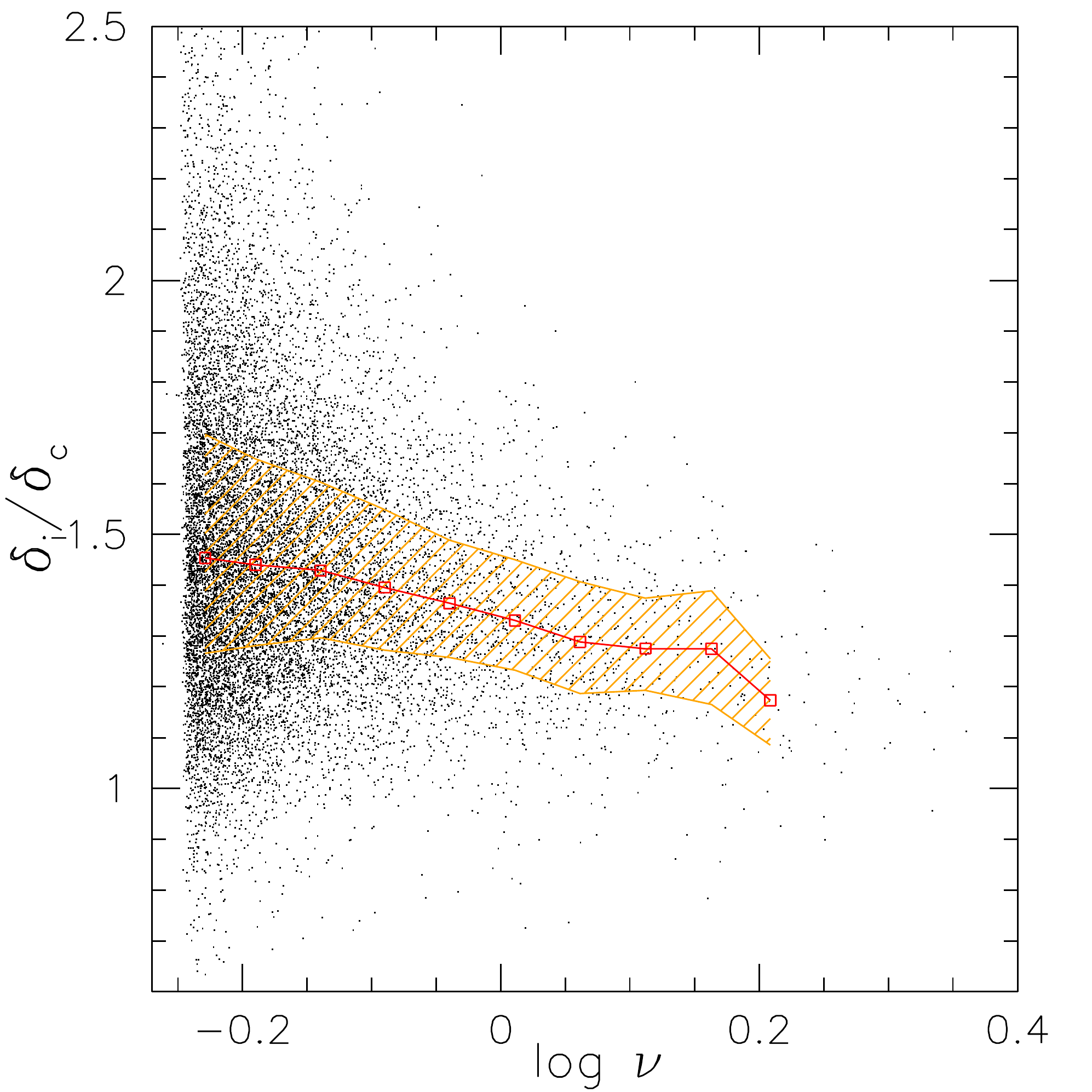}
\caption{Distribution  of  the  initial overdensity as a function 
  of halo mass (here scaled to $\nu$).  
  The overdensity is expressed in units of the critical value in the 
  SC model, for ease of comparison with the EC prediction that it 
  should always be greater than $\delta_{sc}$, and increasingly so  
  at small masses.\label{deltaec}}
\end{figure}

Figure~\ref{deltaec} shows the distribution of $\delta$ as a function 
of halo mass, here scaled to $\nu = \delta_{sc}/\sigma(m)$.  
For ease of comparison  with equation~(\ref{decep}), the overdensity 
within the initial proto-halo has been rescaled to the present time 
using the linear growth factor $D_{+}(z=0)/D_{+}(z=49) = 38.993$.   
It is clear that the initial overdensity is a decreasing function of 
mass, in qualitative agreement with previous work 
\citep{shethmotormen01, robertson09, 2012EliaLudlow}, and with the prediction  which 
comes  from  combining  the  $EC$  model  with  the statistics  of 
Gaussian  random  fields \citep{shethmotormen01}.   The
required overdensity for collapse is  higher for low mass haloes which
must  be  able to  hold  themselves  together  against tidal  effects.
Although the  overdensity values can be substantially  higher than the
critical  value associated  with  the SC  model,  especially at  small
masses, they  are almost never smaller.  This  represents a nontrivial
success of the EC model.

\begin{figure}
 \includegraphics[width=7cm]{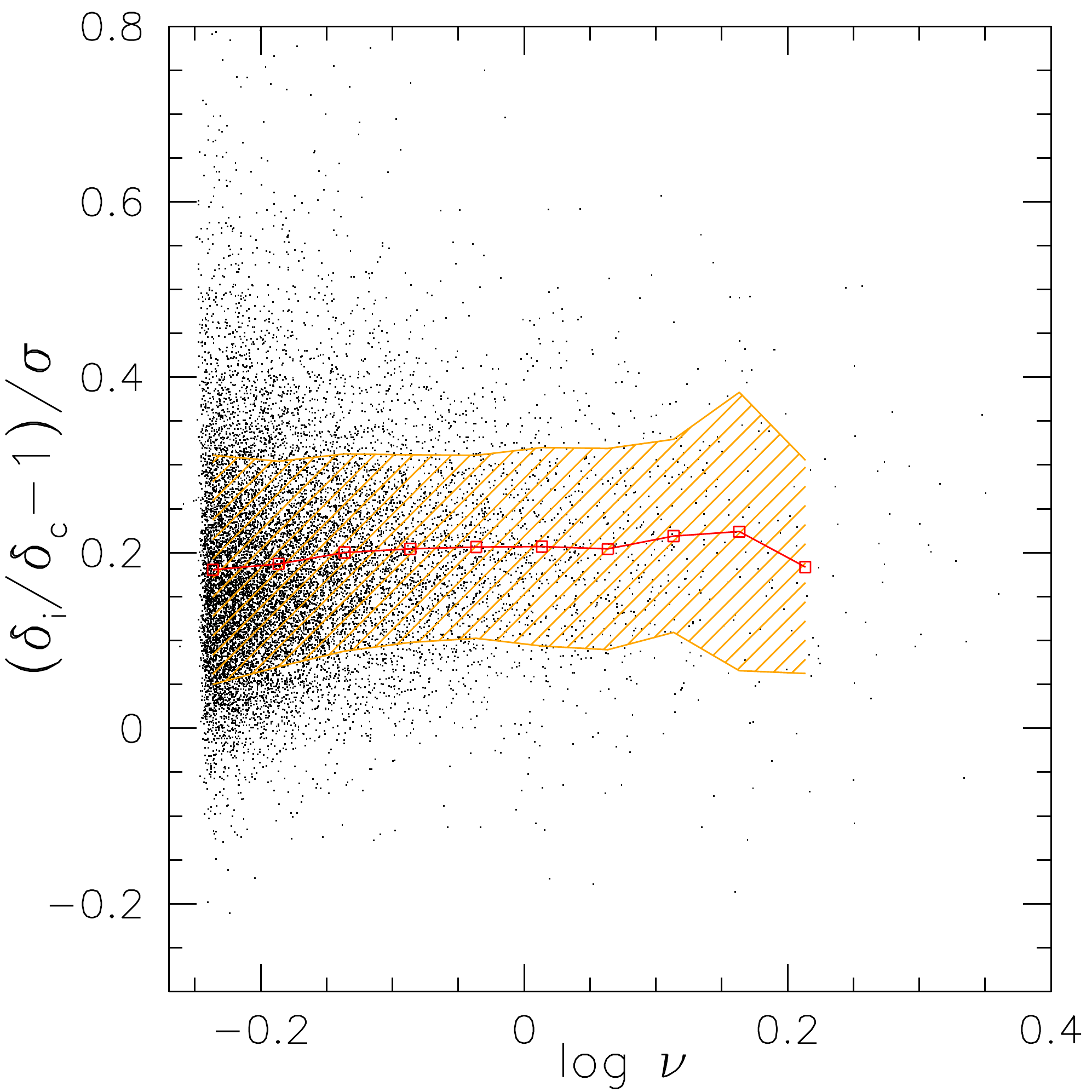}
 \caption{Same as previous figure, except that now the initial overdensity 
  has been scaled to $\Delta_h$, in terms of which most of the mass 
  dependence has been removed.
 \label{deltah}}
\end{figure}

\begin{figure}
\includegraphics[width=7cm]{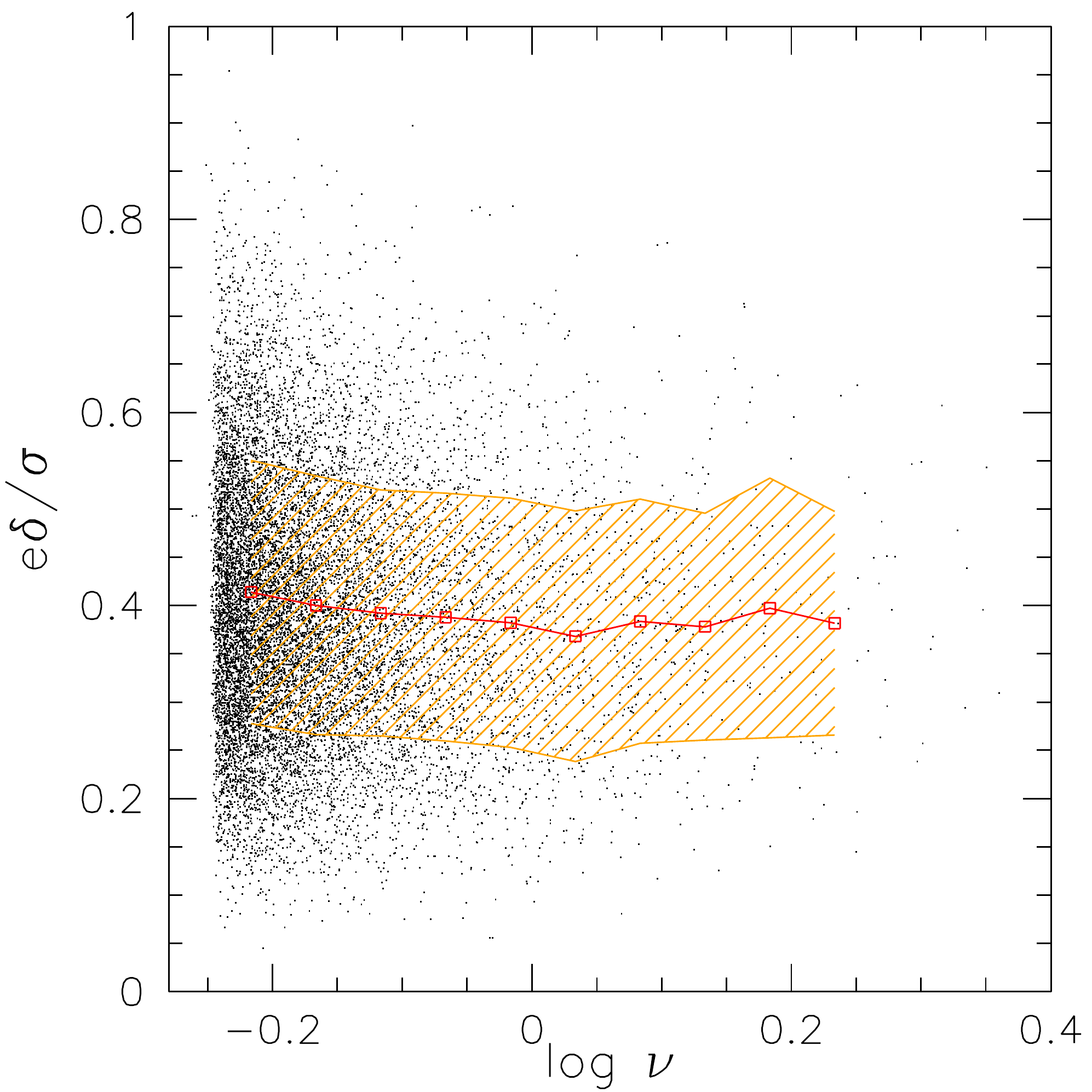}
\includegraphics[width=7cm]{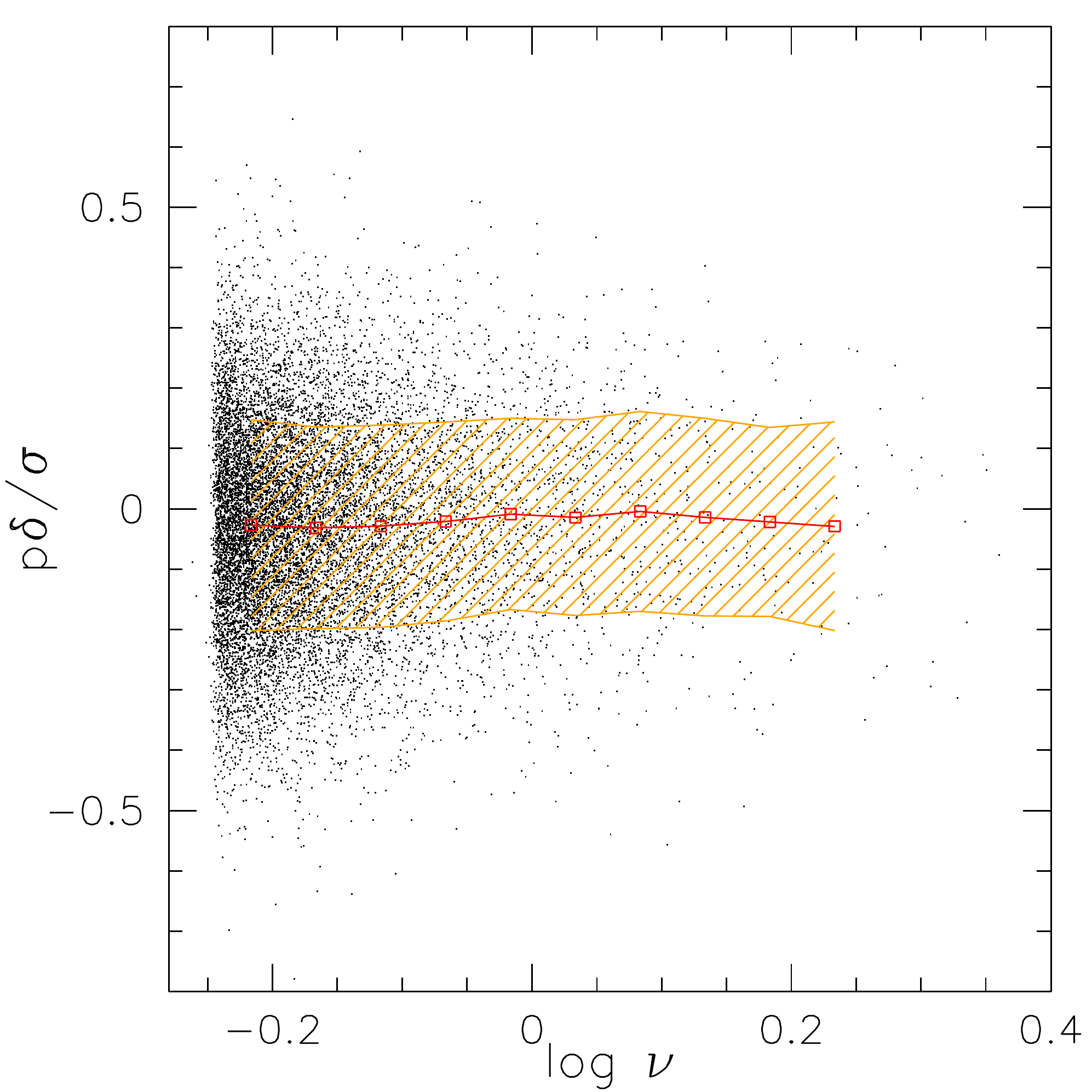}
\caption{Distribution  of  the  initial ellipticity  and
  prolateness of  the haloes as a halo mass (here scaled to $\nu$).  
  $e$ and $p$ have been scaled by $\delta/\sigma$ for 
  more direct comparison with the EC prediction that $e\delta/\sigma$ 
  and $p\delta/\sigma$ should be independent of protohalo mass.\label{inval}}
\end{figure}

Qualitative success does not guarantee quantitative agreement.  
In the EC model, the initial density of protohalo regions depends on 
the shape parametes $e$ and $p$ (equation~\ref{decep}).  
A simple estimate of the mass dependence comes from replacing $e\delta$ 
and $p\delta$ in equation~(\ref{decep}) with a naive estimate of their 
mean values.  If averaged over all possible positions in a Gaussian 
field, this gives $\sigma/\sqrt{5}$ and zero as characteristic 
values of $e\delta$ and $p\delta$, making 
 $\delta/\delta_{sc} = 1 + \beta\,\nu^{-2\gamma} = 1 + 0.25\,\sigma^{1.2}$.  
The median values shown in Figure~\ref{deltaec} are smaller than 
this most naive prediction.  

Most of the mass dependence in Figure~\ref{deltaec} is removed by rescaling to 
 $\Delta_h \equiv (\delta/\delta_{sc} - 1)/\sigma$.
Figure~\ref{deltah} shows that, in these scaled units, the mean and rms 
values are approximately 0.2 and 0.12.  These values are smaller than 
those reported by 
\cite{robertson09} ($0.48/\delta_c = 0.28$ and $0.3/\delta_c = 0.18$).  
Some of this is due to the fact that we use the actual particle 
distribution in the initial conditions to determine $(\delta,e,p)$, 
rather than assuming the initial shape was spherical.  In addition, 
we use EO rather than SO derived quantities.  We believe these 
differences matter, since the shape of the distribution is expected 
to encode information about the quantities in the initial conditions 
which determine halo formation.

\begin{figure*}
\includegraphics[width=18cm]{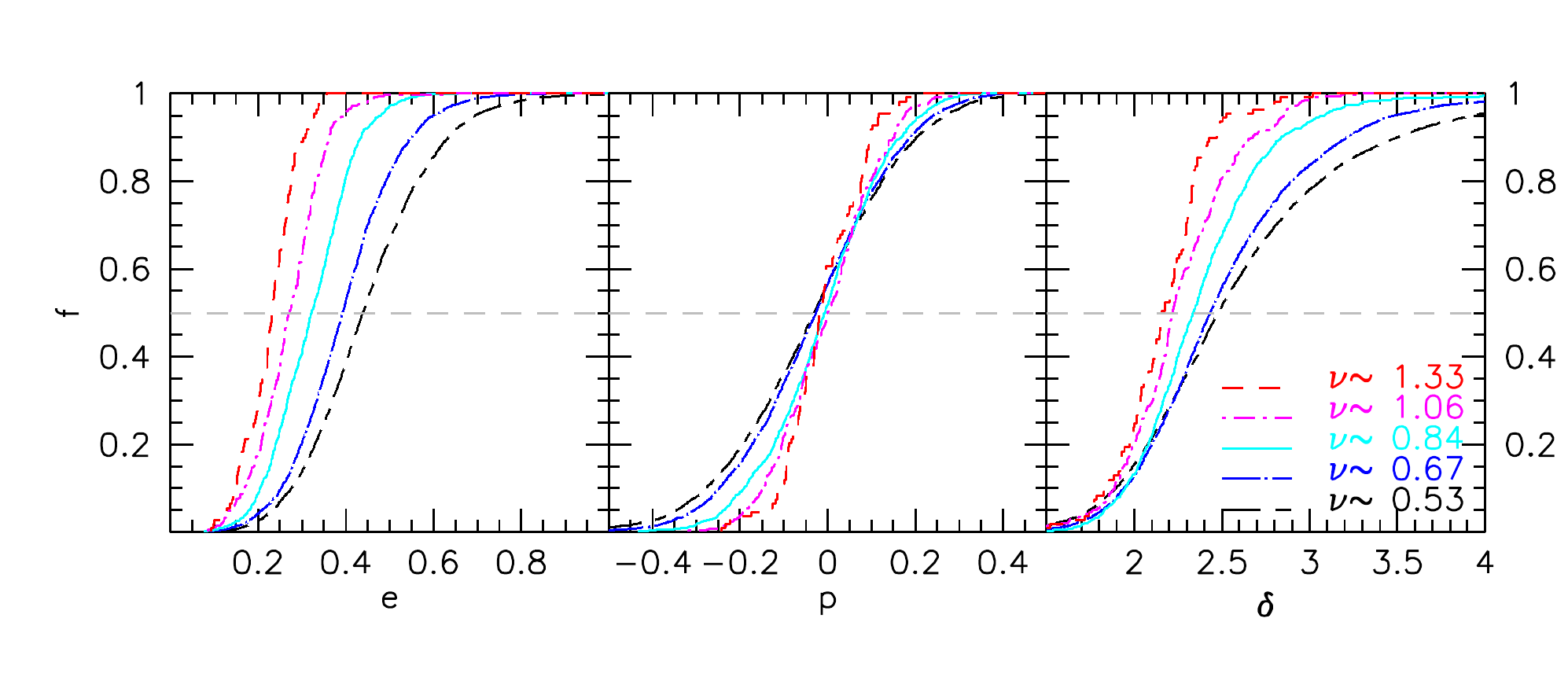}
\caption{Cumulative  distribution of initial ellipticity, prolateness
  and  overdensity, for  five  different mass  bins  indicated by  
  colors and line types;  the  bins have  the  same width  in $\log(\nu)$.   They
  clearly show that ellipticity is a decreasing function of mass, as 
  is the overdensity.  In contrast, prolateness is almost the same, on
  average, for  all mass bins. The  gray dashed line is  drawn to help
  identify the median values in the cumulative distributions. 
  \label{initialdist}}
\end{figure*}

\subsection{Ellipticity and prolateness of the deformation tensor}
In the EC model, the fundamental quantities which influence the 
evolution are not the $\lambda_j$ themselves, but the combinations 
$\delta,e,p$.  Since we have already discussed $\delta$, we now turn 
to a study of $e$ and $p$ in our protohaloes.  
Recall that, if the shape is caused entirely by the 
deformation tensor, then a prolate mass configuration corresponds 
to $p<0$ and an oblate one to $p>0$.  (This classification differs 
from the one given by the  geometrical  mass distribution,  in  
which a physically prolate/oblate halo has $p>0$/$p<0$.)  

The distributions of ellipticity and  prolateness are shown in the two
panels  of  Figure~\ref{inval} as  a  function  of  halo mass  (always
represented by  $\nu$).  Recall that  the combination of the  EC model
with Gaussian field statistics means that lower mass haloes should, on
average, have larger values of $e$ with a larger rms around this mean,
while the most naive  averaging procedure  suggests that $e\delta/\sigma$
should  have  mean  $\approx   1/\sqrt{5}  =  0.447$  and  rms  $0.14$
independent of mass.  The lack of  mass dependence in the mean and rms
values  is in  good agreement  with our  measurements,  although their
actual  values, 0.4  and 0.13,  are slightly  smaller  than predicted.
Similarly,  $p\delta/\sigma$  has  mean  zero as  predicted,  but  the
measured rms of 0.15 is smaller than the predicted value of 0.22.

Inserting these values to obtain the EC prediction for the typical 
overdensity of protohaloes yields 
 $\delta/\delta_{sc} = 1 + \beta\,(4/5)^\gamma \nu^{-2\gamma}
                    = 1 + 0.22\,\sigma^{1.2}$.
Comparison with the bottom panel of Figure~\ref{inval} shows that 
this is about 10\% higher that what we see.  Our measurements 
indicate that $\delta/\delta_{sc} - 1 \approx 0.2 \sigma$, with 
an rms scatter around this mean of $0.12\sigma$ (also see 
Figure~\ref{deltah}).

Before moving on, note that $e\delta \equiv (\lambda_1-\lambda_3)/2$, 
so, on average, $\lambda_1-\lambda_3 \sim 0.8\sigma$.  
In addition, $p=0$ implies $\lambda_1+\lambda_3 = 2\delta/3$ or 
$\lambda_1-\lambda_2 = \lambda_2-\lambda_3$.  While this latter 
is interesting itself, it is also worth noting that the mean values 
we see imply mean values of 
 $\lambda_1 = \delta/3 + 0.4\sigma$,
 $\lambda_2 = \delta/3$ and 
 $\lambda_3 = \delta/3 - 0.4\sigma$.
This shows that the mean value of $\lambda_3$ will be less than zero 
once $\sigma$ exceeds $5\delta/6$.  
Inserting the mean trend $(\delta/\delta_c - 1) \sim 0.2\sigma$ 
implies 
 $\sigma \ge (5\delta_c/6) (1 + 0.2\sigma)$ 
or 
 $\sigma \ge (5\delta_c/6)/(1 - \delta_c/6)$.
So, a significant fraction of haloes with 
$\nu\equiv \delta_c/\sigma \le (6 - \delta_c)/5 \sim 13/15$, 
will tend to have $\lambda_3<0$ (as shown in Figure~\ref{aut_def}).  

Finally, Figure \ref{initialdist} shows the cumulative distributions of 
the three initial  parameters, as a function of halo mass: the  haloes 
were divided into five mass bins, described  in Table \ref{tab_bins}.
This confirms that, on average, both the initial overdensity $\delta$ 
and initial ellipticity $e$ are larger at small mass, while the initial 
prolateness $p$ is distributed around a mean value of zero.

We believe we  have demonstrated that the combination  of the EC model
with the  statistics of the Gaussian potential  field works reasonably
well.   Because the potentials  of the  most massive  proto-haloes are
indeed  more  spherical,  whereas  the  shapes  of  the  most  massive
virialized    haloes    are    less   spherical    (\citet{allgood06};
\citet{schneider12}), we  conclude that we have a  puzzle.  Either the
EC  model is incorrect  in its  description of  the evolution,  or the
proto-haloes are non-spherical even initially, and this influences the
final shapes.

\section{Halo shapes}
The simplest $EC$ model assumes that haloes evolve and collapse 
through a series of triaxial configurations; the directions of the 
three axes of the ellipsoid do not change, and they are determined 
by the initial deformation tensor.  Therefore, the mass tensor is 
perfectly correlated with the initial (Lagrangian space) tidal 
tensor.  In what follows, we first show that the shapes are not 
spherical initially.  This raises the question of whether or not 
the EC assumption that the directions of the principal axes of the 
mass and deformation are aligned is justified. We address this in the 
second half of this section.  

Although we  speak of  the principle  axes of the  two tensors,  it is
clear that  their physical meaning  is different: the  eigenvalues and
eigenvectors  of  the mass  tensor  give  an  estimate of  the  actual
particle distribution  at a given  time, thus describing  the physical
\emph{shape}  of the halo;  whereas those  of the  deformation tensor,
calculated at  the initial time,  describe the characteristics  of the
gravitational potential field within and around protohaloes and so are
used to predict how the protohalo shape will change in time.

\subsection{Triaxiality of initial shapes}
\begin{figure}
\includegraphics[width=8cm]{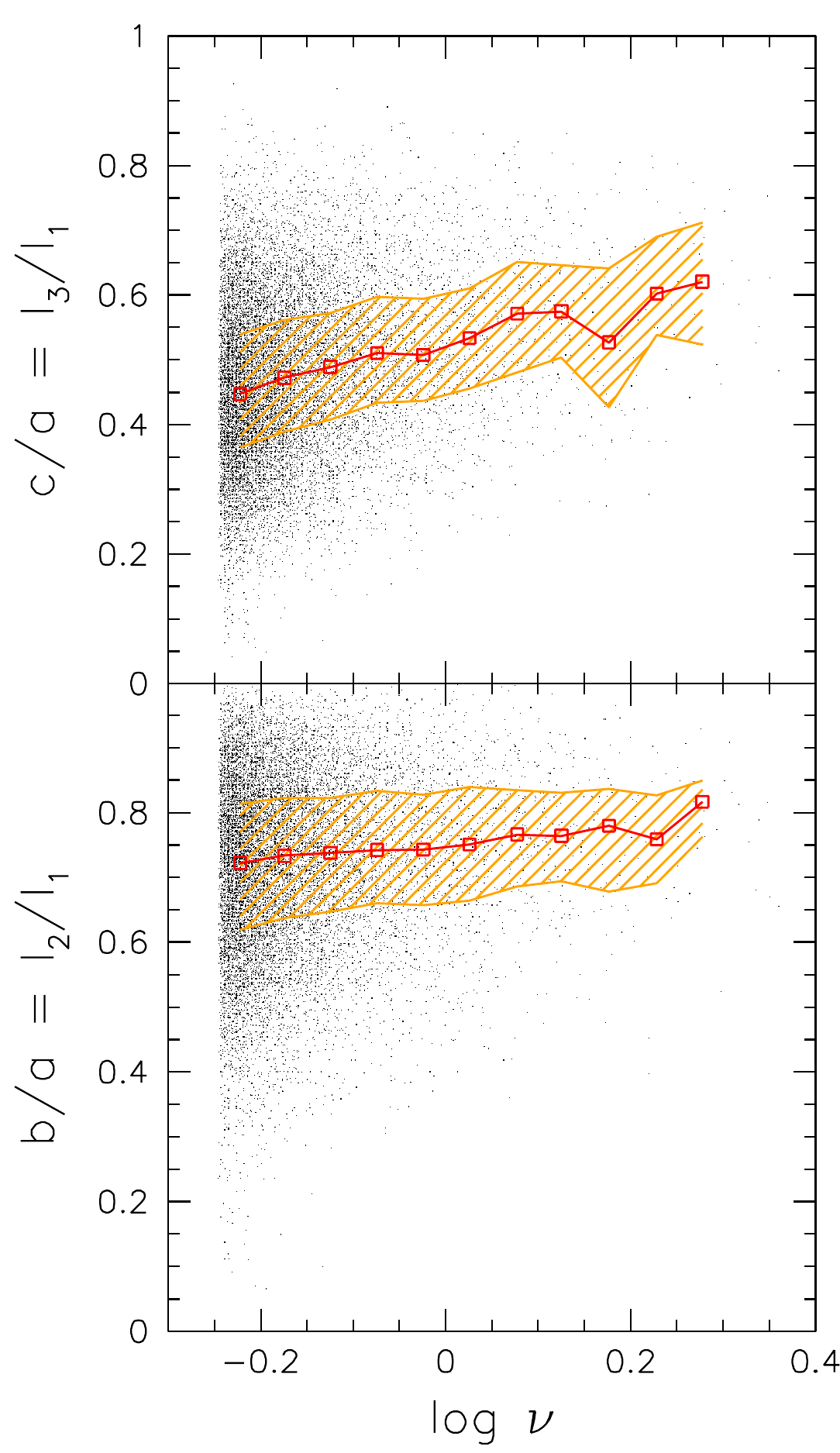}
\caption{ Axis ratios calculated  from the square  root of  the mass
  tensor eigenvalues of the protohaloes at the initial time.\label{ratioLag}}
\end{figure}

Recall that, in the simplest EC model, the initial mass distribution 
was spherical, so the mass tensor at any later time is determined 
completely by the initial tidal tensor defined by its constituent 
particles.  Figure~\ref{ratioLag} shows that the protohaloes are not 
spherical:  the axis ratios as determined from the square-roots 
of the initial mass tensor only approach unity for the most massive 
haloes.  

\subsection{Initial alignment of the mass and deformation tensors}

To quantify the correlation  between the mass and deformation tensors,
we     study     the    the     distribution     of    $\mu_{ij}     =
|\cos(\widehat{\lambda_{i}\l_{j}})|$, the cosine  of the angle between
axis $i$ and $j$ of  the initial deformation and mass tensors.  Recall
that  the  deformation  tensor  eigenvalues  are ordered,  as  in  the
previous Sections,  as $\lambda_{1}\geq\lambda_{2}\geq\lambda_{3}$; we
use  $l_{i}$   ($l_{1}\geq  l_{2}\geq  l_{3}$)  for   the  three  axes
calculated from the  mass tensor (here we prefer  to call them $l_{i}$
instead  of   $a,  b,  c$,   for  uniformity  of  notation   with  the
$\lambda_{i}$s).  The directions are said  to be correlated if they are
well-aligned ($\mu_{ij}\simeq 1$),  while they are $anticorrelated$ in
the  opposite  case  ($\mu_{ij}\simeq  0$): the  maximum  misalignment
happens  when  the two  vectors  (representing  the  semi-axes of  the
ellipsoid) are perpendicular to  each other, while they are considered
aligned when $\cos(\widehat{\lambda_{i}\l_{j}})\simeq  1, -1$, since a
misalignment  of  more than  90  degrees  corresponds  in fact  to  an
alignment on the other side.

\begin{figure}
\includegraphics[width=8cm]{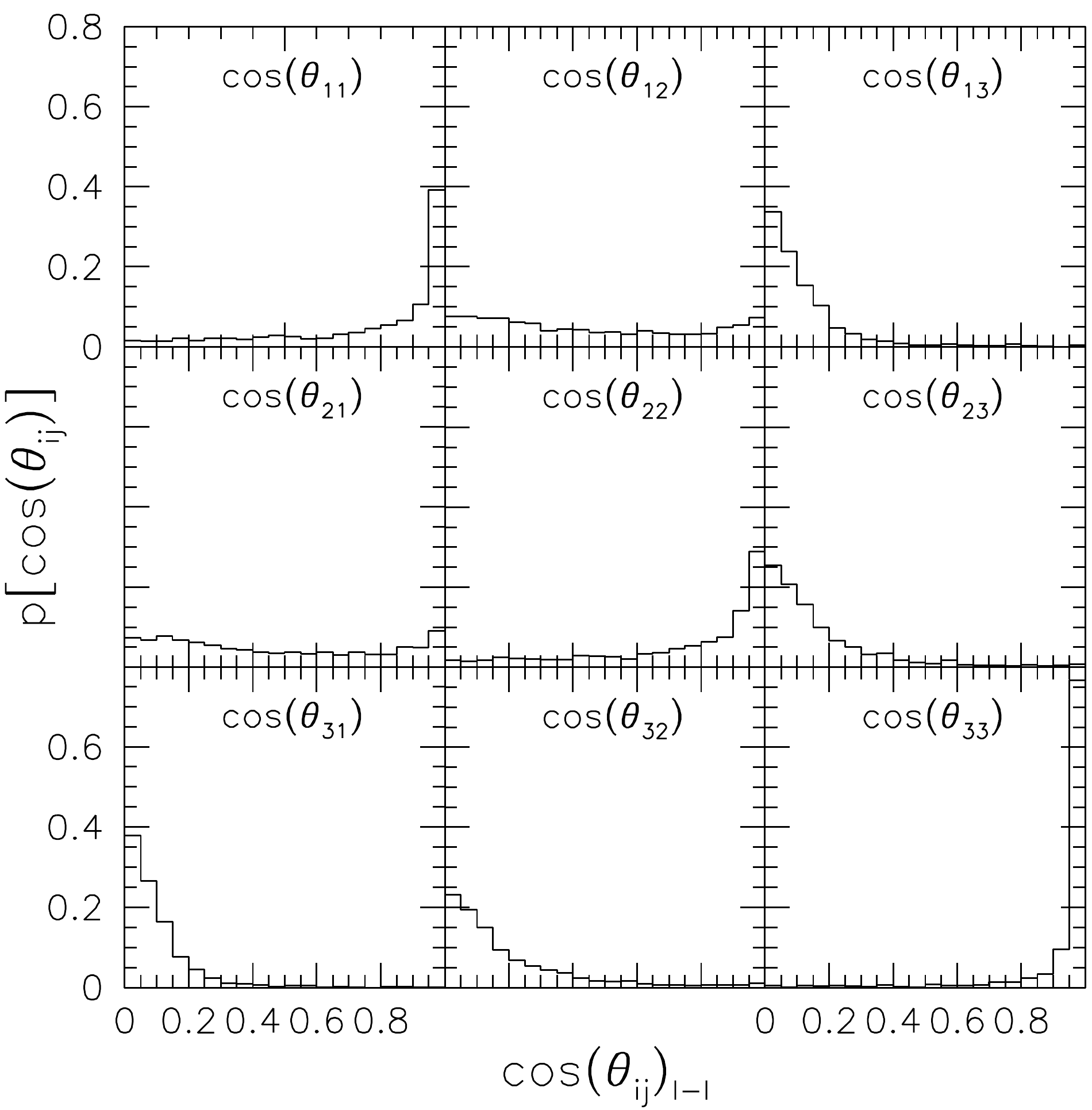}
\caption{Alignment between the principal axes of the initial mass
         tensor and the initial deformation tensor.\label{corr2}}
\end{figure}

Previous work (measurements in  simulations) has shown that while this
correlation is  indeed very  good, the two  tensors are  not perfectly
correlated  \citep{porciani2002b}.  Figure~\ref{corr2}  confirms this.
For our  protohaloes, the longest axis  of the mass  tensor $l_{1}$ is
very well-aligned with $\lambda_{1}$  of the deformation tensor, which
corresponds  to the  direction of  maximum compression;  similarly the
shortest mass  tensor axis  $l_{3}$ is aligned  with the  direction of
minimum  compression  $\lambda_{3}$.   To  quantify this,  the  median
values of the distributions in Figure \ref{corr2} are:
\begin{equation}
p[\cos(\theta_{ij})]=
\begin{pmatrix}
 0.898 &  0.396 & 0.081\\
 0.406 & 0.853 & 0.109 \\
 0.070 & 0.120 & 0.985\\
\end{pmatrix}.
 \label{muIF}
\end{equation}

Thus, it is no longer obvious that the halo will turnaround first 
along the direction of its initial major mass axis, then along the 
second and finally along the third mass axis.  Rather, the differences 
in the turnaround (or collapse) times of the three axes may be smaller 
than they were for the case of collapse from a sphere.  

\begin{figure}
\includegraphics[width=8cm]{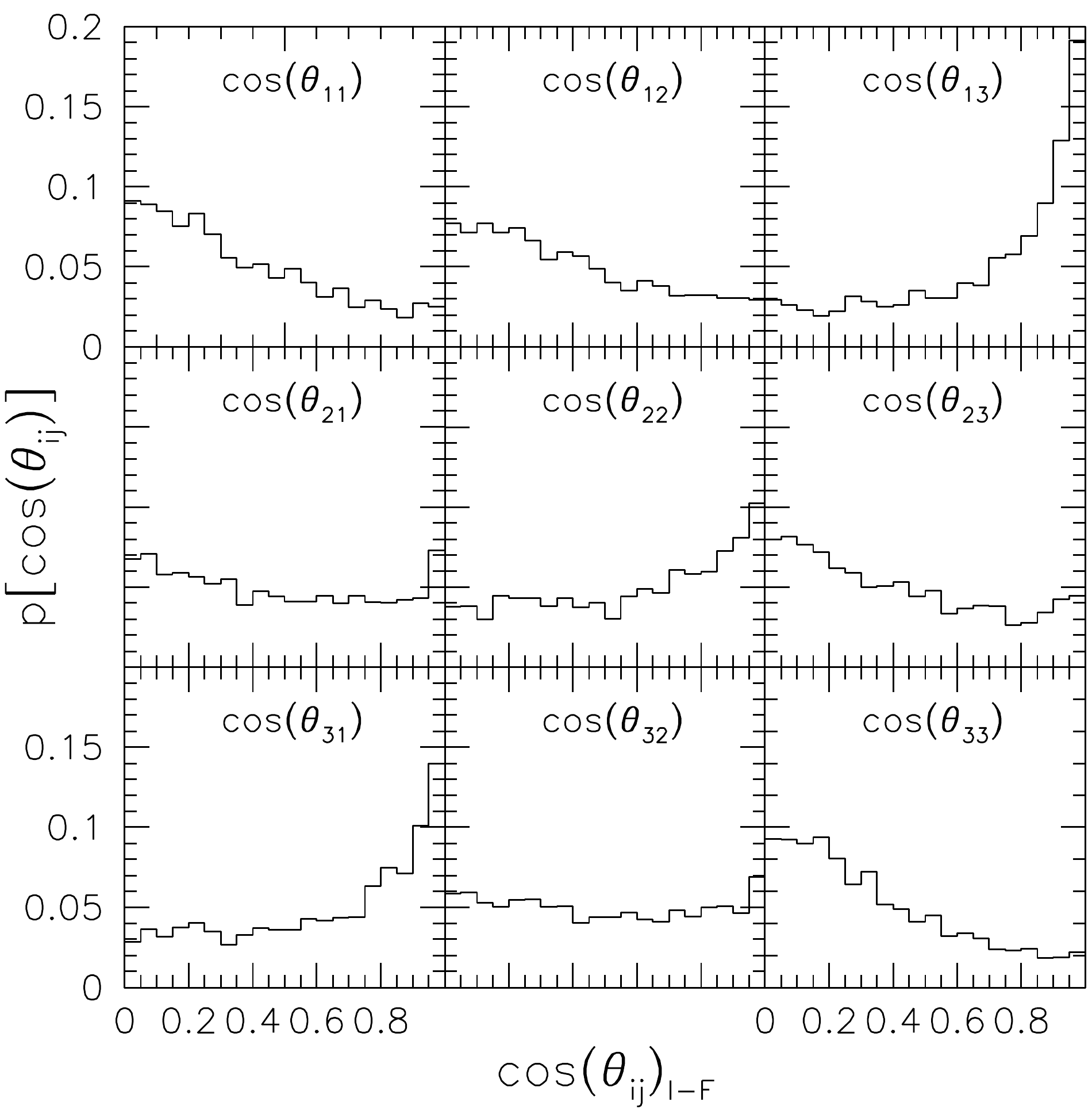}
\caption{Alignment between the principal axes of the  initial
         deformation tensor  and the final mass tensor.\label{corr3}}
\end{figure}

\subsection{Alignment of the final mass tensor with the initial deformation tensors}

Therefore, it is also interesting to study the  alignment between the 
initial and  final axes  to check  how the evolution affects the axis 
orientation.  At the initial time, both the mass and the deformation 
tensor can  be  used   to  approximate protohaloes, while  the final 
orientation and dimension  of the haloes can  be described only by  
the   mass  distribution   (since  the \citet{zeldovich70} approximation 
used  to calculate the $\lambda_{i}$ can be applied  only at the initial 
time).   Hence, Figure \ref{corr3} compares the alignments between the 
principal axes of the final  mass tensor with that of the initial  
deformation tensor.   

For  this particular analysis  we only chose haloes with more than 
1000  particles - so as to ensure accurate determination of  the 
mass eigenvectors - and haloes with smooth evolution and  mass accretion
history.   This last condition arises  because we  have found  that a
significant  fraction  of   haloes  presents  an  irregular  evolution
history, probably  due to the occourence of  important merging events,
which  of course  influence the  final  orientation of  the halo.   We
identified these using an objective automated algorithm, which searches 
for irregularities in the evolution of  the amplitudes of  the axes  of 
the  best fitting  ellipsoid.  We exclude them from  the sample for this 
specific case, because their final properties could be influenced by 
the merging history more than by  the initial distribution, unlike 
regular  haloes.  Thus, they constitute a different population which 
should be studied separately, as we intend to  do in a future work.  

Returning to  Figure \ref{corr3}, it is  clear that at  the final time
the axes of the two tensors  are not so well-aligned as in the initial
conditions:          the          top-right          box,          for
$\cos(\theta_{13})=\cos(\widehat{\lambda_{I1}\l_{F3}})$,   shows  that
the final  shortest axis of the  mass tensor seems to  be aligned with
the  direction  of  initial  maximum compression  $\lambda_{1}$.   The
bottom right  panel shows  that it is  almost anticorrelated  with the
direction  of initial  minimum  compression.  This  sort of  inversion
occurs also in  the case of the first axes:  the bottom-left box shows
that  the final  longest mass  axis  $l_{F1}$ is  better aligned  with
$\lambda_{3}$ than with $\lambda_{1}$ (top-left box).

\begin{figure}
\includegraphics[width=8cm]{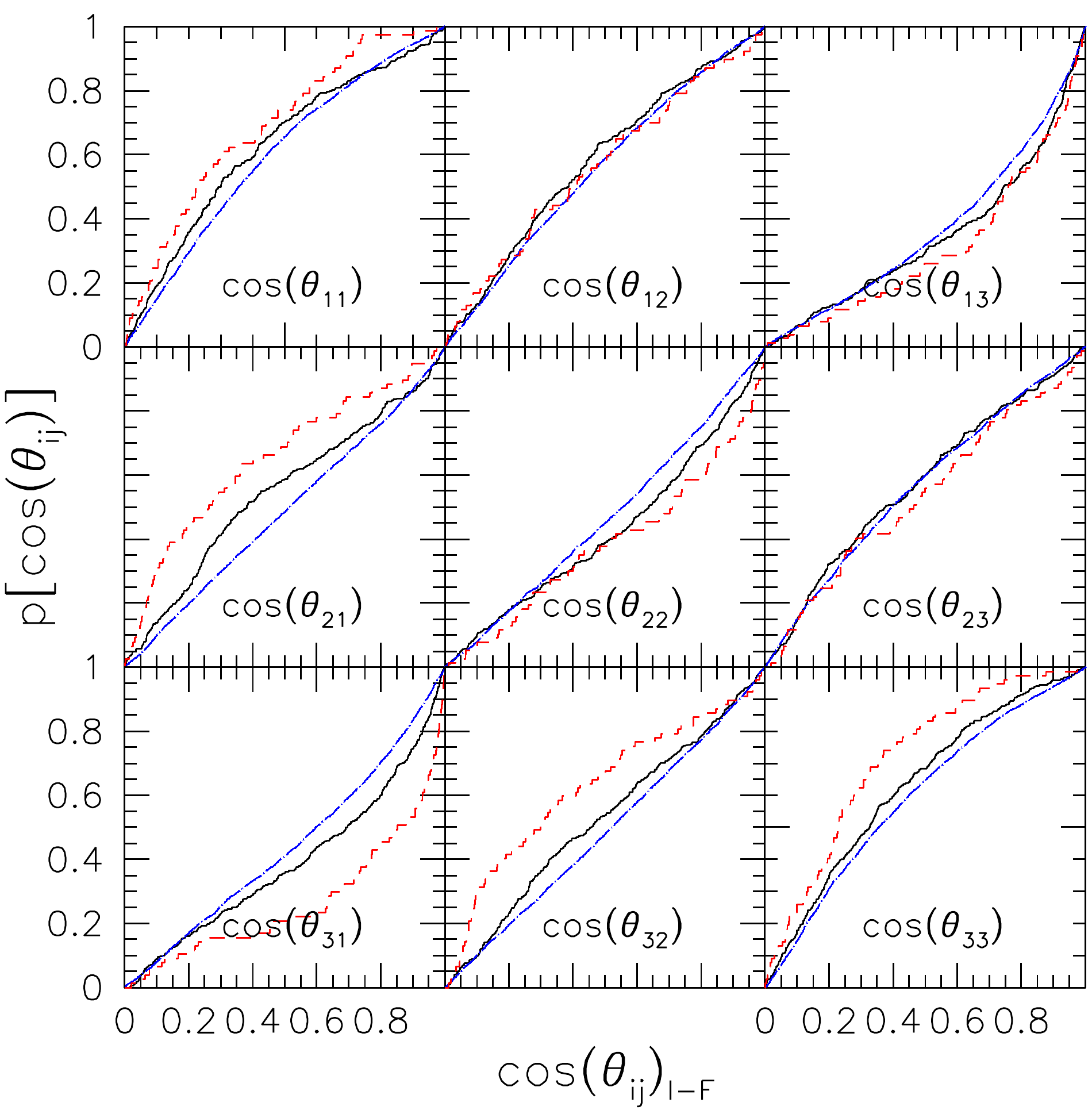}
\caption{Same   distributions  as  in   Figure  \ref{corr3},   but  in
  cumulative  rather  than differential  form;  haloes  are now  split
  according to their  mass: $M_{\star}/16$ in blue (dot  - long dash),
  $M_{\star}$ in black (solid) and $8-16M_{star}$ in red (short dash). The
  alignments    or    misalignments    are    enhanced    at    higher
  masses. \label{corr3bin}}
\end{figure}

The tendency shown in Figure \ref{corr2} is thus completely reversed.  
Now, the median values of the (cosines of the) alignment angles are:
\begin{equation}
p[\cos(\theta_{ij})]=
\begin{pmatrix}
0.310 & 0.360 & 0.779 \\
0.463 & 0.613 & 0.365 \\
0.677 & 0.497 & 0.300 \\
\end{pmatrix}
 \label{muIF}
\end{equation}
confirming that whereas before the 11 and 33 correlations were 
strongest, now it is the 13 and 31 correlations which are strongest.

\subsection{Mass dependence}

Figure~\ref{corr3bin} shows  how this behaviour depends  on halo mass,
plotting  the  same  distributions  of Figure  \ref{corr3}  (but  here
cumulative and not differential) for  haloes of three mass bins: black
(solid),  blue (dot  - long  dash) and  red (short  dash)  curves show
results  for mass  bins  centered on  $M_{\star}$, $M_{\star}/16$  and
$8-16M_{\star}$.   Clearly,   the  evolution  pattern   that  we  have
suggested is strongest for the most massive haloes.

We also  checked the alignment between  the initial and  final axes of
the  mass tensor,  but we  do not  present the  result here  since the
correlations are weaker and only  the behavior of the shortest axis is
well defined:  this seems  to indicate that  the potential field  is a
better  tracer of  the initial  shape, since it shows what will be the
evolution tendency more than the actual initial position of particles.

\subsection{Evolution}
\begin{figure}
\includegraphics[width=8cm]{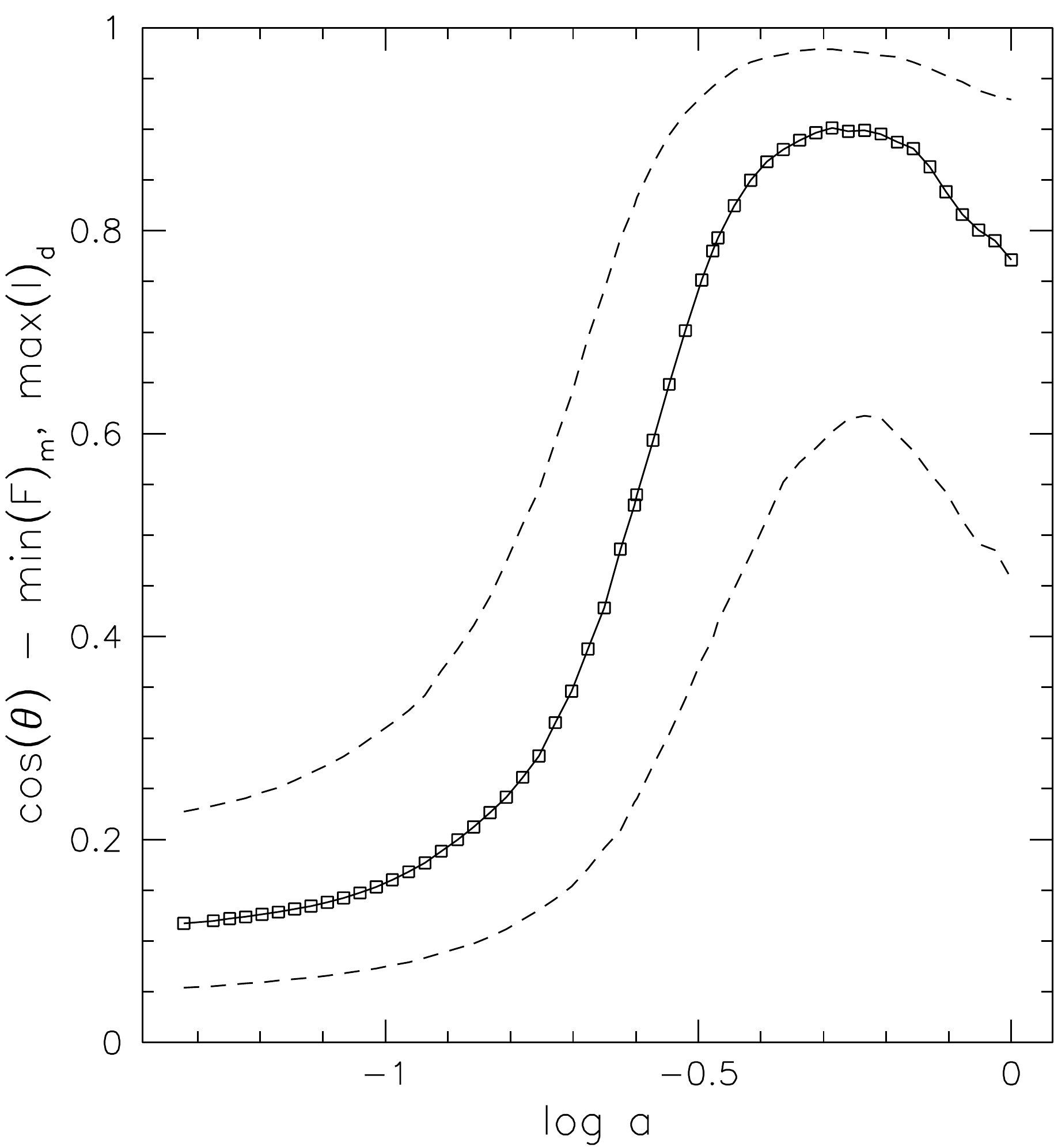}
\caption{Misalignment between the  longest axis of initial deformation
  tensor and the the shortest axis of the mass tensor of the haloes at
  as a  funcion of time,  represented by log(a): the  points represent
  the median  values at each time  and the dashed lines  the first and
  third quartiles of the distribution.\label{align_first}}
\end{figure}

To  study   more  closely  if   and  when  the  axes   invert,  Figure
\ref{align_first} shows the evolution  of the misalignment between the
major  axis  of  initial  deformation tensor  $\lambda_{I1}$  and  the
shortest  axis  of  the  mass  tensor  $l_{3}$  (i.e.,  the  strongest
alignment shown  in equation~\ref{muIF}) calculated at  each time step
of the simulation.  The points show  the median value of the cosine of
the angle between  the principal axes of the two  tensors at each time
and he dashed lines the first and third quartiles of the distribution.
At high redshift the two  axes are almost perpendicular to each other,
but  by $z=0$  they  are  almost perfectly  aligned.   I.e., the  halo
collapses in the direction of maximum compression and, by the end, the
shortest  axis of the  halo lies  in this  direction (which  is almost
perpendicular to  the direction of the initial  shortest axis).  There
is, in fact, an intermediate period of rapid misalignment, followed by
a stable period at late times. While Figure \ref{align_first} shows
the median behaviour, some examples of the evolution of individual
haloes can be found in Appendix \ref{appendix1}.  

We believe that  this inversion in direction is  rather generic to the
collapse process,  but there are  a number of  ways in which  this can
happen.  If  the mass and  deformation tensors are  perfectly aligned,
then this  is just a consequence  of the compression due  to the tidal
field.  I.e., early-on, the protohalo contracts most rapidly along the
direction of  $\lambda_{1}$, while  contracting more slowly  along the
direction of $\lambda_{3}$.  Eventually, the protohalo collapses first
along  the direction  of $\lambda_{1}$,  then along  $\lambda_{2}$ and
finally along  $\lambda_{3}$; the resulting mass  distribution ends up
being  more   compressed  along  the  direction   of  maximum  initial
compression,  $\lambda_{1}$, and  more elongated  in the  direction of
initial minimum  compression $\lambda_{3}$.  As a  result, the longest
axis of the mass tensor of the final object, which is perpendicular to
the  shortest  axis,  ends  up  being  perpendicular  to  its  initial
direction,  even though  there is  no rotation.   In this  case, there
should be a time during the evolution when the axis lengths are equal.

If the mass and deformation tensors are not well-aligned initially, 
then the object may rotate.  For example, if the second axes are 
well-aligned but the first and third are not, then the object may 
rotate about the second axis.  This rotation will be most effective 
if the sum of the first and third deformation vectors is perpendicular 
to the sum of the first and third mass vectors.  I.e., rotation will 
be most efficient if $\lambda_1$ and $\lambda_3$ have opposite signs.  
Once the axes are aligned, the rotation stops.  In this case, the 
exchange of direction of the axes need not be accompanied by an 
exchange of lengths.  Note that although we showed that $\mu_{11}\sim 1$, 
suggesting that the initial misaligments are small, $\mu_{11}=0.9$ 
still allows misalignments as large as $20^\circ$.  Since the rotation 
depend on the cross- rather than dot-product of the two vectors, 
rotation might be more common than one might have concluded on the 
basis of the statement that $\mu_{11}\sim 1$.  

Finally, there could be an  intermediate case: it may be that the
halo  does not  actually rotate,  in the  sense of  having  an overall
angular momentum:  rather, as  the particle distribution is squeezed by
different amounts in different directions, the relative lengths of the
mass axes change, the particle distribution deforms (maybe not exactly
along  the  directions  of  the  $\lambda_{i}$s  due  to  the  initial
misalignment between  the two tensors) and so the direction  in which
the longest axis points can evolve.

In summary: the major axis of a protohalo is initially aligned with the 
direction of maximal compression \citep[in agreement with]{porciani2002b}.  
For a few haloes, the compression is sufficiently large that this 
longest axis ends up being the first one to turnaround and collapse, 
and so it becomes the shortest axis.  Whether or not the axis lengths 
invert, for most haloes, the axis directions do:  the shortest axis of 
the final object ends up being well-aligned with the initial direction 
of maximal compression.  As can be argued from the examples in
Appendix \ref{appendix1}, haloes show a general behaviour, but they
differ in the details of the evolution: further analyses are needed to
provide a more complete description of the process.

\section{Summary and Conclusions}\label{discuss}
To study the shape  and evolution properties  of dark matter haloes, 
we  analysed a sample  of simulated haloes extracted from the GIF2 
simulation \citep{gao04}. First, we showed how the way in which 
haloes are identified affects conclusions about their shapes, and 
of the proto-halo patches from which they formed.  

In Section~3 we presented an algorithm which identified dark matter 
haloes  as  triaxial  ellipsoids enclosing the virial overdensity 
assumed by theory.  In Section~4 we showed that:
\begin{itemize}
\item Compared to the usual spherical overdensity method, haloes 
  identified using our  \emph{Ellipsoidal Overdensity} method are 
  more massive, because a triaxial shape is able to trace the 
  density distribution better than a sphere.  However, this 
  difference is typically less than 10\%.  
\item The change to the final halo shape is more  significant, with 
  differences in the axial ratios as large as 40\%, leading to  more 
  elliptical (and in particular more prolate) shapes: 
  our \emph{Ellipsoidal Overdensity} criterion refines halo shapes, 
  making them more varied and realistic. 
\end{itemize}
In Section~5 we studied the corresponding protohaloes in the initial 
conditions. We checked how the EO algorithm affects protohalo 
properties, showing that the eigenvalues of the deformation tensor 
have a slightly broader distribution,  but otherwise exhibit the 
same trends as for protohaloes of SO haloes. This suggests that the 
initial potential field is less affected by the selection algorithm. 

In the EC model, the properties of the initial field are expressed 
not in terms of the three eigenvalue of the deformation tensor 
$\lambda_{i}$ but using the trace, and the shape parameters $e$ and $p$.
We plotted their distributions, finding that:
\begin{itemize}
 \item for most protohaloes the eigenvalues of the deformation tensor 
  do not have the same sign; the fraction of protohaloes for which this 
  is true increases at low masses; 
 \item at fixed mass, protohaloes have a range of initial overdensities;  
  these are almost always larger than the critical overdensity associated
  with spherical collapse; the mean overdensity increases as mass decreases, 
  scaling approximately as $\delta_c(1+0.2\sigma)$,  and the rms distribution 
  around this mean is $0.2\sigma$ (it is broader for lower mass haloes); 
 \item the median ellipticity $e$ of the deformation tensor decreases 
  as mass increases:  the distribution of $e\delta/\sigma$ is approximately
  independent of halo mass, having mean $\sim 0.4$ and rms 0.14;
 \item the median prolateness of $p\delta/\sigma\simeq 0$ with rms $0.15$;
 \item the mass tensors are increasingly non-spherical as protohalo 
  mass decreases.
\end{itemize}
The middle three findings are in good qualitative agreement with the  
triaxial collapse model in which low mass haloes need a higher initial 
overdensity so as to collapse by the present time, because they tend 
to be less spherical.  

The final part of this work aimed at understanding how the initial
potential field interacts with  haloes and influences their evolution,
since the initial  properties are the dominant ingredient  in the $EC$
model. Thus, we studied the cross-talk between the mass tensor and the
deformation tensor: the first gives an estimate of the particle
distribution of haloes and so of their actual shape and orientation,
while the second (calculated only at the initial time) describes the
potential field.  We showed that:
\begin{itemize}
\item At the initial time, the principal axes of the two tensors are 
  very well-aligned; the longest axis of the mass tensor ($l_{1}$) 
  is aligned with the direction of maximum compression $\lambda_{1}$,
  $l_{2}$ is aligned with $\lambda_{2}$, and $l_{3}$ with $\lambda_{3}$.
  However, although $\cos$ of the misalignment angle is $\sim 1$, the 
  angle itself can still be of order tens of degrees.   
\item At the final time ($z=0$) the alignment between the axes is 
  reversed, as a consequence of the collapse process and the
  associated deformation of the particle distribution.
\item The change in directions of the first and third axes are sometimes
  dominated by the different compression factors, and others by what 
  appears to be rotation, although it is still unclear if the apparent 
  rotation is actually an asymmetrical deformation of the particle 
  distribution. 
\end{itemize}

Whereas the initial alignment is  common in models where  haloes form
from peaks in  the initial density field, the  order of the alignments
is opposite to that for peaks, which predict alignments which are more
like those we see for final haloes \citep{vandeWeygaert96}.

These late time alignments, which suggest that the direction of 
maximal compression remains approximately constant as the halo 
collapses and its shape changes, are consistent with recent 
measurements of anisotropies in the halo-mass clustering signal, which 
indicate that the long axes of virialized haloes are aligned with the 
large scale structures in which they are embedded \citep{2012falt}.  
Moreover, a model which explicitly assumes this (lack of evolution of 
the deformation tensor) to be true appears to provide a good description 
of these measurements \citep{2012papaisheth}.  
We plan to study further whether it is deformation or rotation which 
is the cause of the change in alignments we see.

\section*{Acknowledgments}
GD was funded by PhD grants at the University of Padova. 
RKS was supported in part by NSF-0908241 and NASA NNX11A125G.

\bibliographystyle{mn2e}
\bibliography{HaloShapes}

\begin{thebibliography}{}

\bibitem[\protect\citeauthoryear{{Allgood}, {Flores}, {Primack}, {Kravtsov},
  {Wechsler}, {Faltenbacher} \& {Bullock}}{{Allgood} et~al.}{2006}]{allgood06}
{Allgood} B.,  {Flores} R.~A.,  {Primack} J.~R.,  {Kravtsov} A.~V.,  {Wechsler}
  R.~H.,  {Faltenbacher} A.,    {Bullock} J.~S.,  2006, \mnras, 367, 1781

\bibitem[\protect\citeauthoryear{{Bailin} \& {Steinmetz}}{{Bailin} \&
  {Steinmetz}}{2004}]{bailin04}
{Bailin} J.,  {Steinmetz} M.,  2004, \apj, 616, 27

\bibitem[\protect\citeauthoryear{{Bailin} \& {Steinmetz}}{{Bailin} \&
  {Steinmetz}}{2005}]{bailin05}
{Bailin} J.,  {Steinmetz} M.,  2005, \apj, 627, 647

\bibitem[\protect\citeauthoryear{{Bett}, {Eke}, {Frenk}, {Jenkins}, {Helly} \&
  {Navarro}}{{Bett} et~al.}{2007}]{bett07}
{Bett} P.,  {Eke} V.,  {Frenk} C.~S.,  {Jenkins} A.,  {Helly} J.,    {Navarro}
  J.,  2007, \mnras, 376, 215

\bibitem[\protect\citeauthoryear{{Bond} \& {Myers}}{{Bond} \&
  {Myers}}{1996}]{bondmyers96}
{Bond} J.~R.,  {Myers} S.~T.,  1996, ApJS, 103, 1

\bibitem[\protect\citeauthoryear{{Desjacques}}{{Desjacques}}{2008}]{desjacques08}
{Desjacques} V.,  2008, \mnras, 388, 638

\bibitem[\protect\citeauthoryear{{Eke}, {Cole} \& {Frenk}}{{Eke}
  et~al.}{1996}]{eke1996}
{Eke} V.~R.,  {Cole} S.,    {Frenk} C.~S.,  1996, \mnras, 282, 263

\bibitem[\protect\citeauthoryear{{Elia}, {Ludlow} \& {Porciani}}{{Elia}
  et~al.}{2012}]{2012EliaLudlow}
{Elia} A.,  {Ludlow} A.~D.,    {Porciani} C.,  2012, \mnras, 421, 3472

\bibitem[\protect\citeauthoryear{{Faltenbacher}, {Li} \& {Wang}}{{Faltenbacher}
  et~al.}{2012}]{2012falt}
{Faltenbacher} A.,  {Li} C.,    {Wang} J.,  2012, \apjl, 751, L2

\bibitem[\protect\citeauthoryear{{Gao}, {White}, {Jenkins}, {Stoehr} \&
  {Springel}}{{Gao} et~al.}{2004}]{gao04}
{Gao} L.,  {White} S.~D.~M.,  {Jenkins} A.,  {Stoehr} F.,    {Springel} V.,
  2004, \mnras, 355, 819

\bibitem[\protect\citeauthoryear{{Giocoli}, {Moreno}, {Sheth} \&
  {Tormen}}{{Giocoli} et~al.}{2007}]{giocoli07a}
{Giocoli} C.,  {Moreno} J.,  {Sheth} R.~K.,    {Tormen} G.,  2007, \mnras, 376,
  977

\bibitem[\protect\citeauthoryear{{Giocoli}, {Tormen} \& {van den
  Bosch}}{{Giocoli} et~al.}{2008}]{giocoli08b}
{Giocoli} C.,  {Tormen} G.,    {van den Bosch} F.~C.,  2008, \mnras, 386, 2135

\bibitem[\protect\citeauthoryear{{Gunn}}{{Gunn}}{1977}]{gunn77}
{Gunn} J.~E.,  1977, \apj, 218, 592

\bibitem[\protect\citeauthoryear{{Jing} \& {Suto}}{{Jing} \&
  {Suto}}{2002}]{jingsuto02}
{Jing} Y.~P.,  {Suto} Y.,  2002, \apj, 574, 538

\bibitem[\protect\citeauthoryear{{Kauffmann} \& {White}}{{Kauffmann} \&
  {White}}{1993}]{kauffwhite93}
{Kauffmann} G.,  {White} S.~D.~M.,  1993, \mnras, 261, 921

\bibitem[\protect\citeauthoryear{{Lacey} \& {Cole}}{{Lacey} \&
  {Cole}}{1993}]{lacey93}
{Lacey} C.,  {Cole} S.,  1993, \mnras, 262, 627

\bibitem[\protect\citeauthoryear{{Lacey} \& {Cole}}{{Lacey} \&
  {Cole}}{1994}]{lacey94}
{Lacey} C.,  {Cole} S.,  1994, \mnras, 271, 676

\bibitem[\protect\citeauthoryear{{Lam} \& {Sheth}}{{Lam} \&
  {Sheth}}{2008}]{lamsheth08}
{Lam} T.~Y.,  {Sheth} R.~K.,  2008, \mnras, 389, 1249

\bibitem[\protect\citeauthoryear{{Lee} \& {Shandarin}}{{Lee} \&
  {Shandarin}}{1998}]{leeshan98}
{Lee} J.,  {Shandarin} S.~F.,  1998, \apj, 500, 14

\bibitem[\protect\citeauthoryear{{P{\'a}pai} \& {Sheth}}{{P{\'a}pai} \&
  {Sheth}}{2013}]{2012papaisheth}
{P{\'a}pai} P.,  {Sheth} R.~K.,  2013, \mnras, 429, 1133

\bibitem[\protect\citeauthoryear{{Porciani}, {Dekel} \& {Hoffman}}{{Porciani}
  et~al.}{2002}]{porciani2002b}
{Porciani} C.,  {Dekel} A.,    {Hoffman} Y.,  2002, \mnras, 332, 339

\bibitem[\protect\citeauthoryear{{Robertson}, {Kravtsov}, {Tinker} \&
  {Zentner}}{{Robertson} et~al.}{2009}]{robertson09}
{Robertson} B.~E.,  {Kravtsov} A.~V.,  {Tinker} J.,    {Zentner} A.~R.,  2009,
  \apj, 696, 636

\bibitem[\protect\citeauthoryear{{Rossi}, {Sheth} \& {Tormen}}{{Rossi}
  et~al.}{2011}]{rossi10}
{Rossi} G.,  {Sheth} R.~K.,    {Tormen} G.,  2011, \mnras, 416, 248

\bibitem[\protect\citeauthoryear{{Schneider}, {Frenk} \& {Cole}}{{Schneider}
  et~al.}{2012}]{schneider12}
{Schneider} M.~D.,  {Frenk} C.~S.,    {Cole} S.,  2012, \jcap, 5, 30

\bibitem[\protect\citeauthoryear{{Seljak} \& {Zaldarriaga}}{{Seljak} \&
  {Zaldarriaga}}{1996}]{cmbfast1996}
{Seljak} U.,  {Zaldarriaga} M.,  1996, \apj, 469, 437

\bibitem[\protect\citeauthoryear{{Shen}, {Abel}, {Mo} \& {Sheth}}{{Shen}
  et~al.}{2006}]{shen06}
{Shen} J.,  {Abel} T.,  {Mo} H.~J.,    {Sheth} R.~K.,  2006, \apj, 645, 783

\bibitem[\protect\citeauthoryear{{Sheth}, {Mo} \& {Tormen}}{{Sheth}
  et~al.}{2001}]{shethmotormen01}
{Sheth} R.~K.,  {Mo} H.~J.,    {Tormen} G.,  2001, \mnras, 323, 1

\bibitem[\protect\citeauthoryear{{Sheth} \& {Tormen}}{{Sheth} \&
  {Tormen}}{2002}]{shethtormen02}
{Sheth} R.~K.,  {Tormen} G.,  2002, \mnras, 329, 61

\bibitem[\protect\citeauthoryear{{Springel}, {White}, {Tormen} \&
  {Kauffmann}}{{Springel} et~al.}{2001b}]{springel01b}
{Springel} V.,  {White} S.~D.~M.,  {Tormen} G.,    {Kauffmann} G.,  2001b,
  \mnras, 328, 726

\bibitem[\protect\citeauthoryear{{Tormen}, {Moscardini} \& {Yoshida}}{{Tormen}
  et~al.}{2004}]{tormen04}
{Tormen} G.,  {Moscardini} L.,    {Yoshida} N.,  2004, \mnras, 350, 1397

\bibitem[\protect\citeauthoryear{{van de Weygaert} \& {Bertschinger}}{{van de
  Weygaert} \& {Bertschinger}}{1996}]{vandeWeygaert96}
{van de Weygaert} R.,  {Bertschinger} E.,  1996, \mnras, 281, 84

\bibitem[\protect\citeauthoryear{{Vera-Ciro}, {Sales}, {Helmi}, {Frenk},
  {Navarro}, {Springel}, {Vogelsberger} \& {White}}{{Vera-Ciro}
  et~al.}{2011}]{Vera-CiroAq11}
{Vera-Ciro} C.~A.,  {Sales} L.~V.,  {Helmi} A.,  {Frenk} C.~S.,  {Navarro}
  J.~F.,  {Springel} V.,  {Vogelsberger} M.,    {White} S.~D.~M.,  2011,
  \mnras, 416, 1377

\bibitem[\protect\citeauthoryear{{Warren}, {Quinn}, {Salmon} \&
  {Zurek}}{{Warren} et~al.}{1992}]{warren92}
{Warren} M.~S.,  {Quinn} P.~J.,  {Salmon} J.~K.,    {Zurek} W.~H.,  1992, \apj,
  399, 405

\bibitem[\protect\citeauthoryear{{White}}{{White}}{1996}]{white1996}
{White} S.~D.~M.,  1996, in {Schaeffer} R.,  {Silk} J.,  {Spiro} M.,
  {Zinn-Justin} J.,  eds, Cosmology and Large Scale Structure {Formation and
  Evolution of Galaxies}.
p.~349

\bibitem[\protect\citeauthoryear{{White} \& {Silk}}{{White} \&
  {Silk}}{1979}]{whitesilk1979}
{White} S.~D.~M.,  {Silk} J.,  1979, \apj, 231, 1

\bibitem[\protect\citeauthoryear{{Zel'dovich}}{{Zel'dovich}}{1970}]{zeldovich70}
{Zel'dovich} Y.~B.,  1970, \aap, 5, 84

\end{thebibliography}

\appendix 
\section{Evolution of the particle distribution}\label{appendix1}
The main text described the mean  behavior of the halo population.  
This mean behaviour is indeed representative of individual haloes.  
To illustrate this, we show how the shape and orientation of the 
particle distribution in a few representative haloes evolves.  

Since the deformation and mass tensors are not aligned in general, 
the objects may rotate as well as deform, so we must make some choices 
about how we describe the evolution.  
We have chosen to show two dimensional projections of the particle 
distributions with respect to a fixed coordinate system.  In addition, 
we show how the (square-roots of the) eigenvalues of the mass tensor 
evolve.  These represent the lengths of the principal axes of the 
object; if two of these cross, then this signals that the compression 
due to the deformation tensor has managed to change the relative axis 
lengths.  The main text argued that this is generic in the Zeldovich 
approximation with perfect alignment.
This information about how the size of the object changes gives no 
insight into the spatial orientation of the object.  To see if the principal 
axes of the mass tensor change direction -- from the combined effects 
of compression and rotation -- we also show how the angle between the 
mass tensor axis $l_i$ and the initial deformation tensor axis 
$\lambda_i$ evolves.  

\begin{figure*}
\includegraphics[width=14cm]{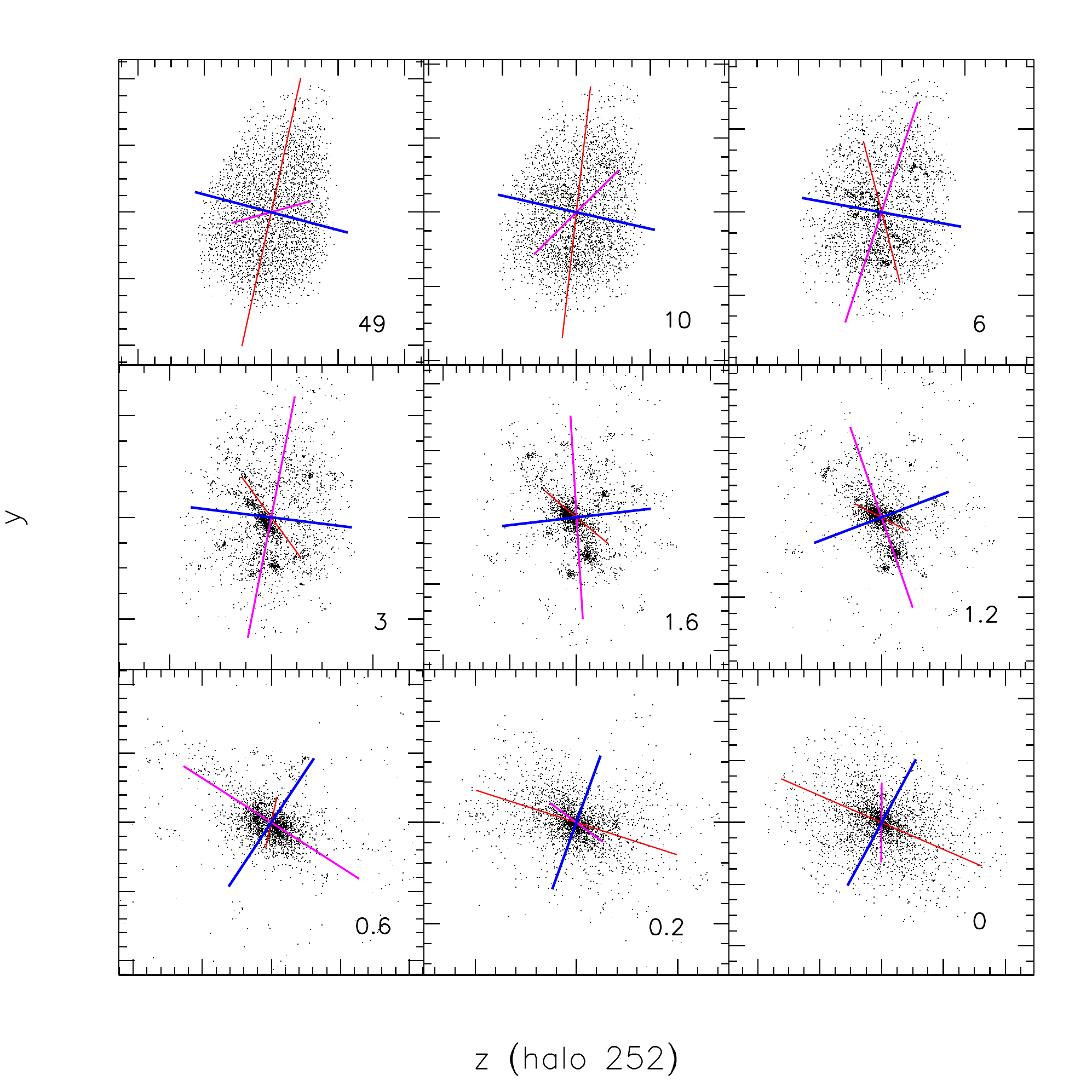}
\includegraphics[width=8cm]{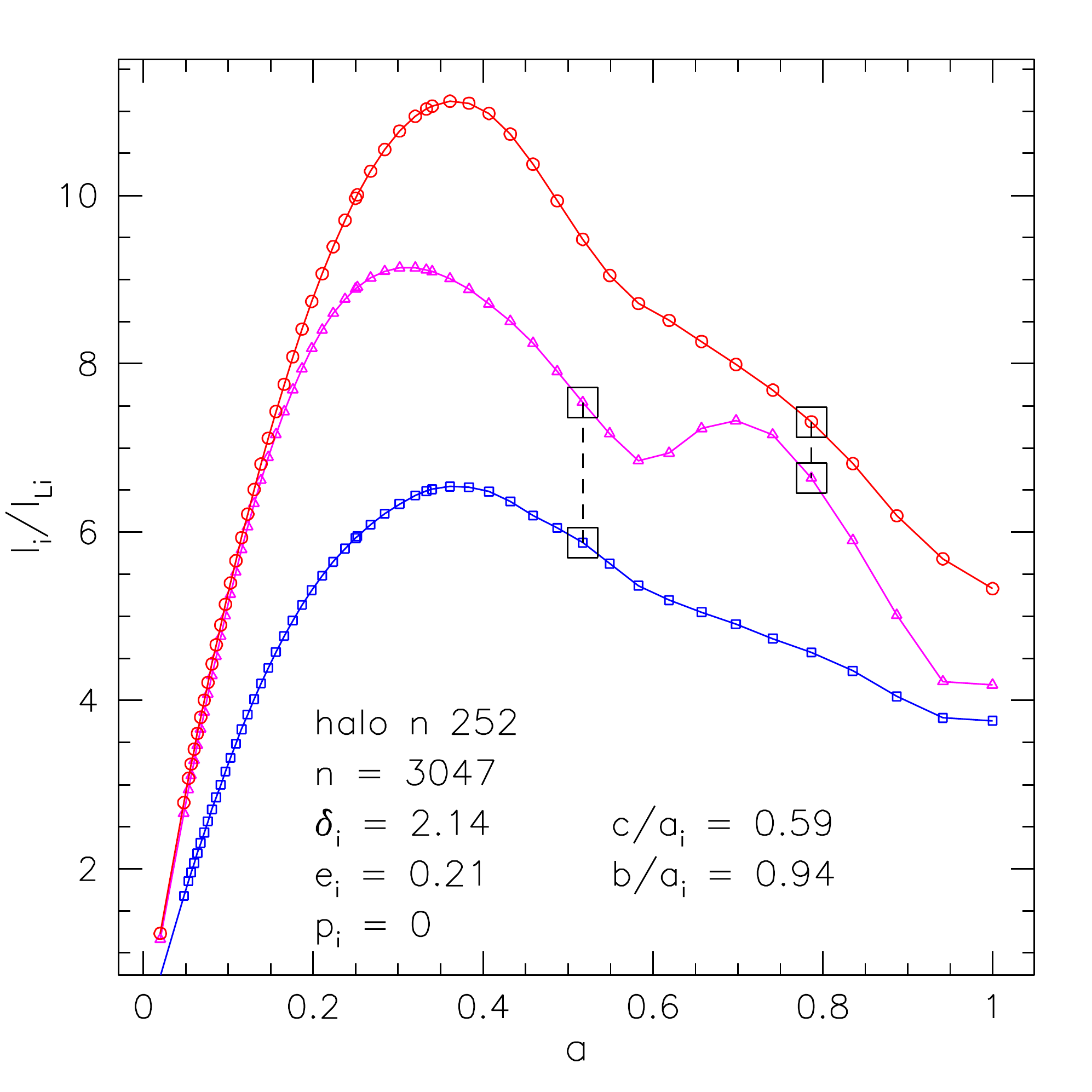}
\includegraphics[width=8cm]{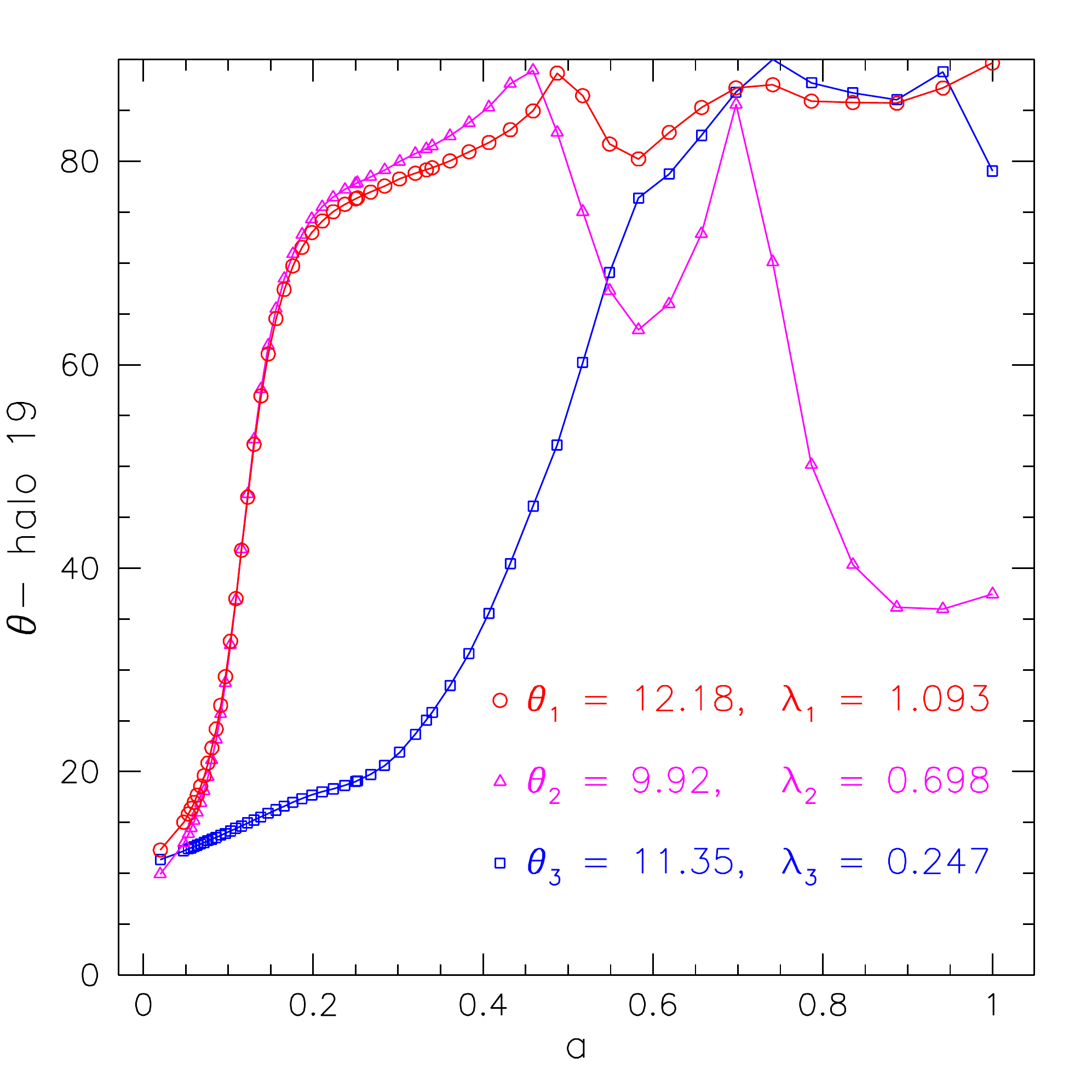}
\caption{Evolution of an object of mass $M_*$.   The axis color scheme is the same
  in all the panels: blue - shortest, magenta - medium, red - longest. \textbf{Top}: 
  the projected particle distribution at 9 different redshifts; the
  thinnest red line stands for the longest axis, while the think blue
  one for the shortest axis at each time step.  \textbf{Bottom left}:  the evolution of the three
  mass axes; the black squares show the moments of inversion of
  direction between two axes.  \textbf{Bottom right}: the evolution of the angle between the mass 
  tensor axes and the initial deformation tensor axes
  ($\widehat{l_{1}\lambda_{1}}$ - red, $\widehat{l_{2}\lambda_{2}}$ -
  magenta, $\widehat{l_{3}\lambda_{3}}$ - blue, with the same point
  kinds of the left panel).\label{haloM*}}
\end{figure*}

\begin{figure*}
\includegraphics[width=15cm]{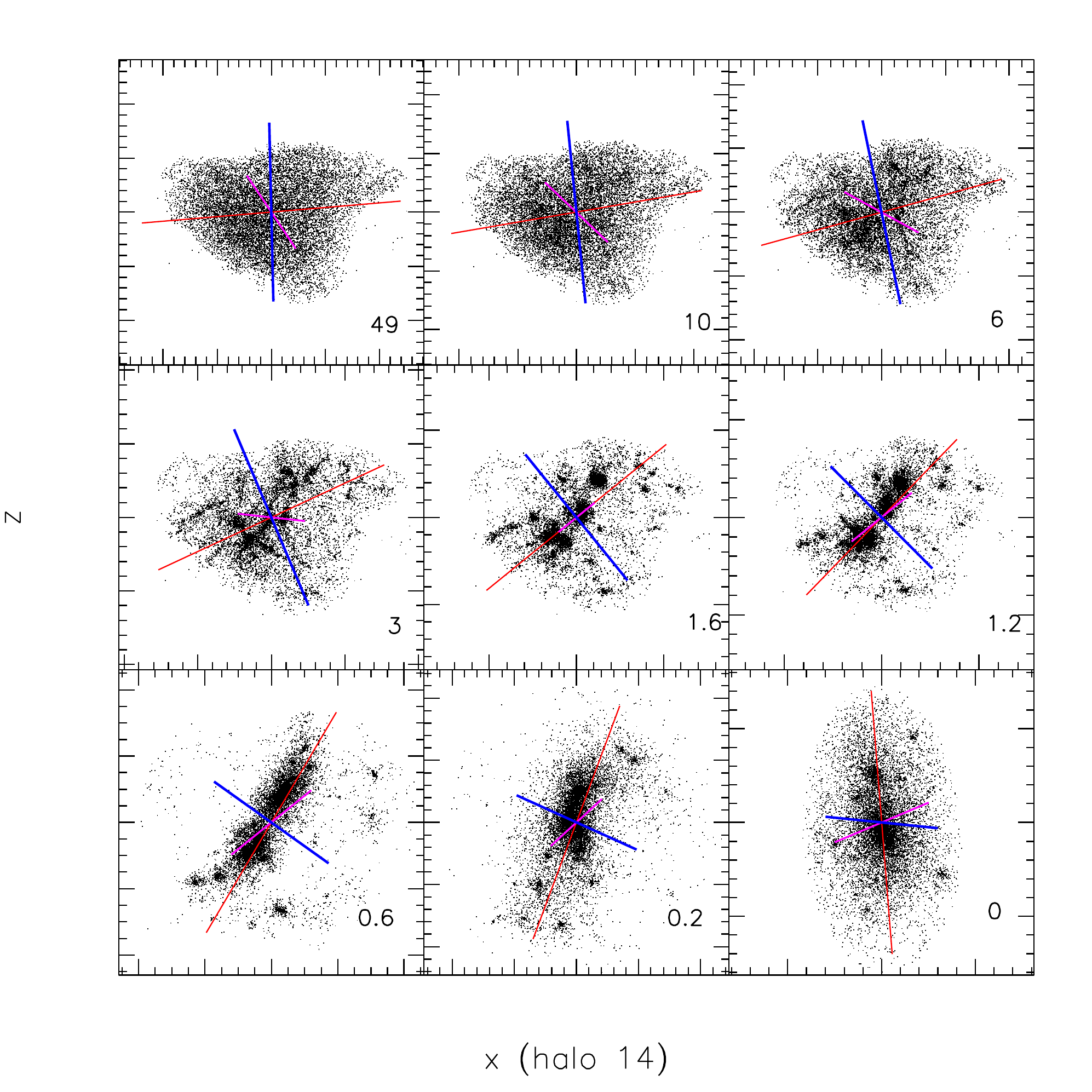}
\includegraphics[width=8cm]{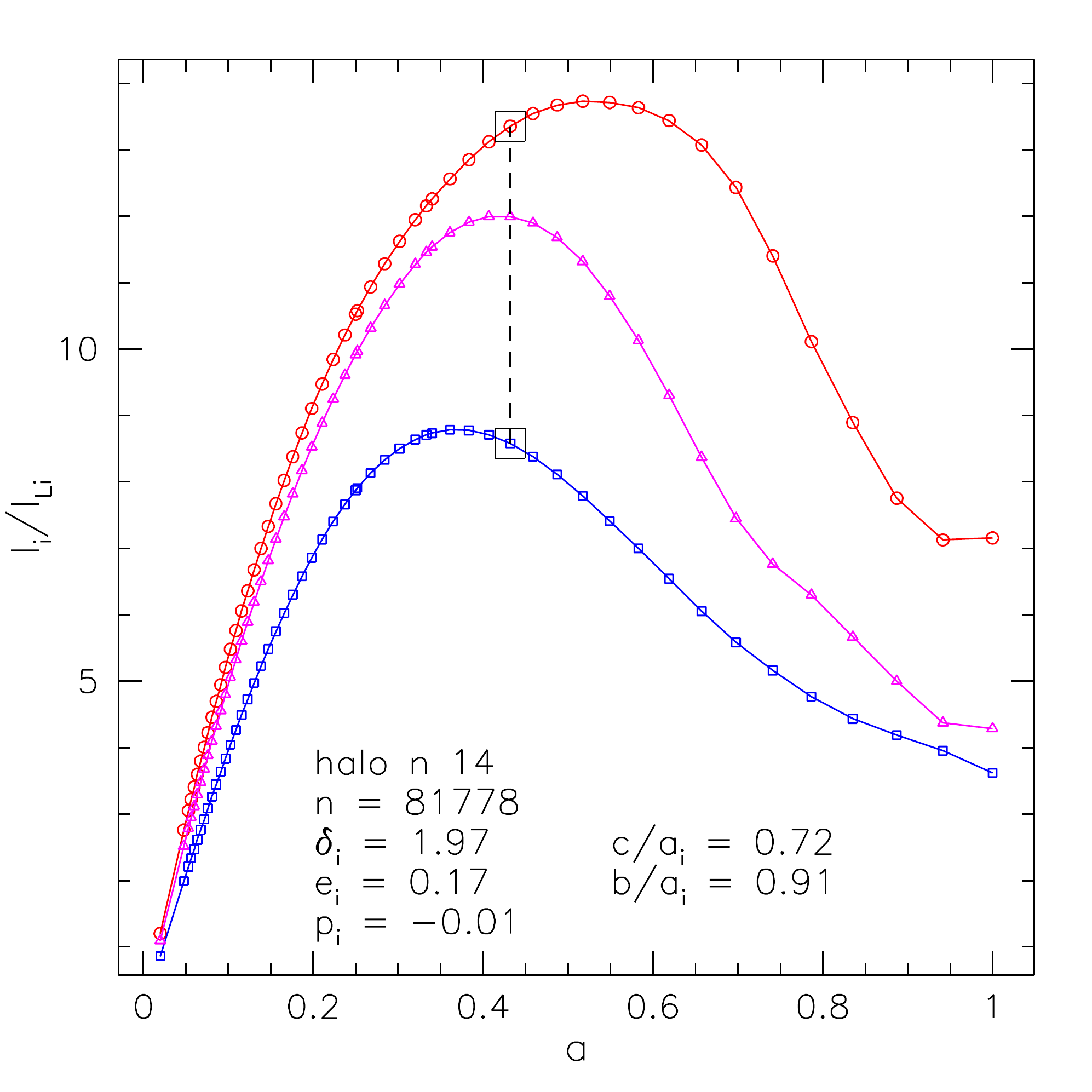}
\includegraphics[width=8cm]{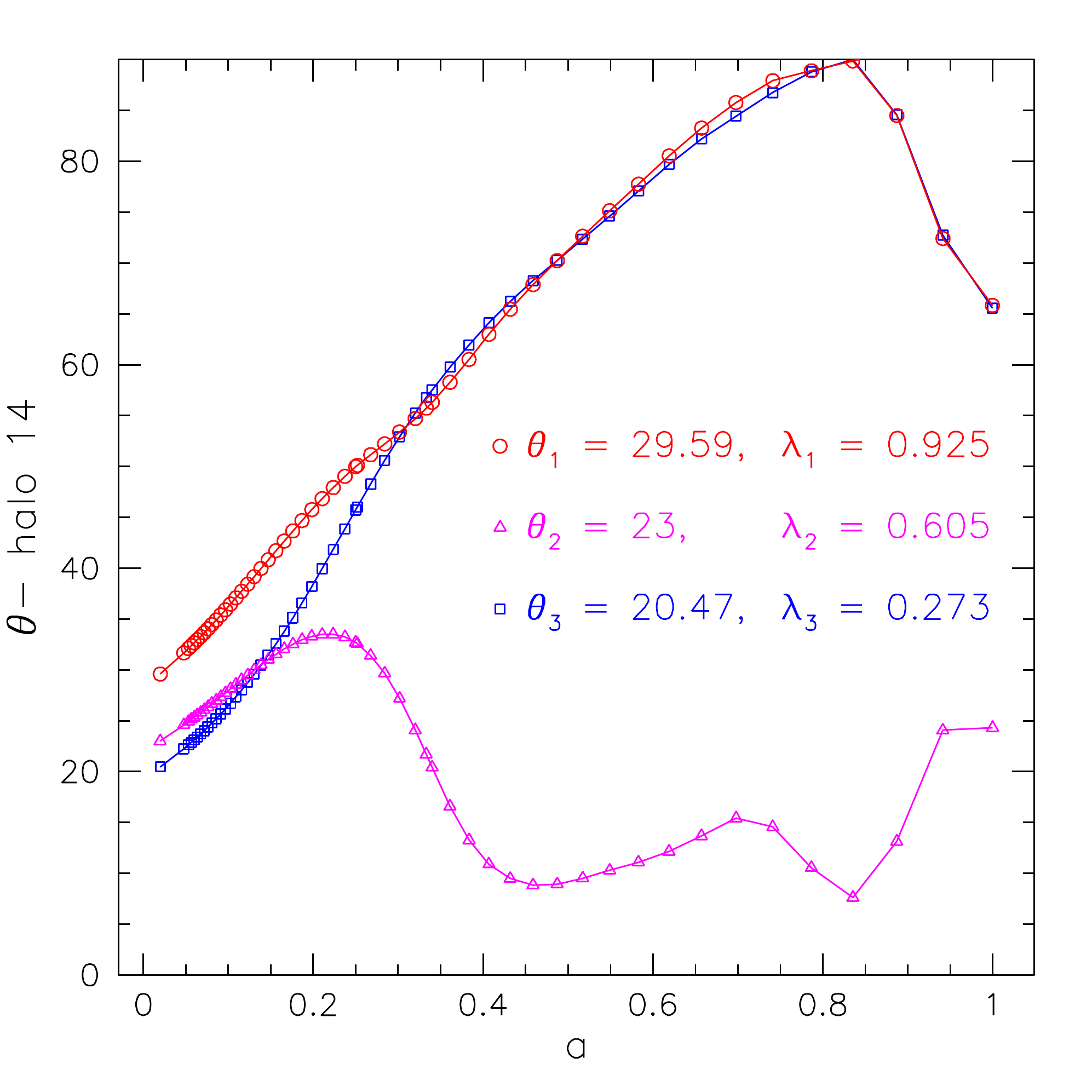}
\caption{Same as previous figure, but now for an object of mass $16M*$.
         \label{halo16M*}}
\end{figure*}

Figures~\ref{haloM*} and ~\ref{halo16M*} show the evolution of 
haloes of mass $M_*$and $16M_*$ respectively.  Despite the order 
of magnitude difference in mass, both objects evolve rather 
similarly.  In both cases, 
 the initial particle distribution is rather non-spherical, after 
 which gravitational collapse occurs along the preferred directions 
 as discussed in the main text, creating a pancake; 
 the directions of the three principal axes of the best-fitting 
 ellipsoid change with time;
 there is an \emph{axis inversion} feature, such that the 
 longest and shortest axes exchange directions.  
Notice that the evolution is not identical, even though the values 
of $\delta\sim 2$, $e\sim 0.2$ and $p\sim 0$ are approximately the 
same.  This shows that the initial deformation tensor does not uniquely 
determine the evolution -- the initial shape also matters, as does the 
degree of initial misalignment.  For these haloes, $b/a$ and $c/a$ were 
$(0.94,0.59)$ and $(0.91,0.72)$, respectively, and although the 
initial alignments are all within 10$^\circ$ for the first object, 
they are much worse for the second.  

In  more   detail,  the  top   panels  show  the   projected  particle
distribution  at  nine time  steps  between  $z=49$  and $z=0$.   Red,
magenta and blue  show the longest, intermediate and  shortest axes of
the mass  tensor (with increasing  line thickness from the  longest to
the shortest).  Projecting from three to two dimensions means that the
relative lengths are not always obvious  in such a plot, so the bottom
left panels show how the lengths  of the three mass axes evolve.  Red,
magenta  and   blue  curves  show   the  evolution  of   the  longest,
intermediate  and shortest axes,  in units  of the  initial lagrangian
radius:
\begin{equation}
 r_{{\rm L}i} \equiv (l_{1i} l_{2i} l_{3i})^{1/3}.
\end{equation}

In some cases (but not these), the axis which was initially the longest 
may become the second longest; this sort of length-inversion occurs in 
only a few of our haloes (which were  excluded from the analysis 
regarding the axes  alignments, to  have a  more homogeneous sample).  
The time at which the axes lengths cross does not necessarily coincide 
with the time that the long axis starts becoming better aligned with 
the direction of minimum initial compression -- squares connected by 
dashed lines show the time at which the exchange in direction occurs.  

The bottom right panels illustrate the evolution of the alignment 
angle in more detail.  They show the angle between the mass axes and 
the direction of the corresponding initial deformation axis.  
Red, magenta and blue show the longest, middle and shortest axes
($\widehat{l_{1}\lambda_{i1}}$, $\widehat{l_{2}\lambda_{i2}}$ and 
$\widehat{l_{3}\lambda_{i3}}$, respectively).  Thus, 
Figure~\ref{haloM*} shows that halo 252 appears to rotate about 
its shortest axis initially; by the time the other two axes have 
exchanged directions, they have also reached turnaround.  Thereafter, 
the object rotates about its longest axis, until the other two 
have approximately exchanged directions.  The net result is that 
the longest and shortest axes have exchanged directions: the moment in
which these inversions occur are also labelled in the left panel by
the black squares.

In contrast, the more massive halo 14 shown in Figure~\ref{halo16M*} 
is slightly simpler.  The initial misalignment in this case was larger, 
but its second axis soon aligns with the intermediate axis of the 
initial deformation tensor, after which the object appears to rotate 
about this second axis until the first and third axes have exchanged 
directions.  

\begin{figure*}
\includegraphics[width=15cm]{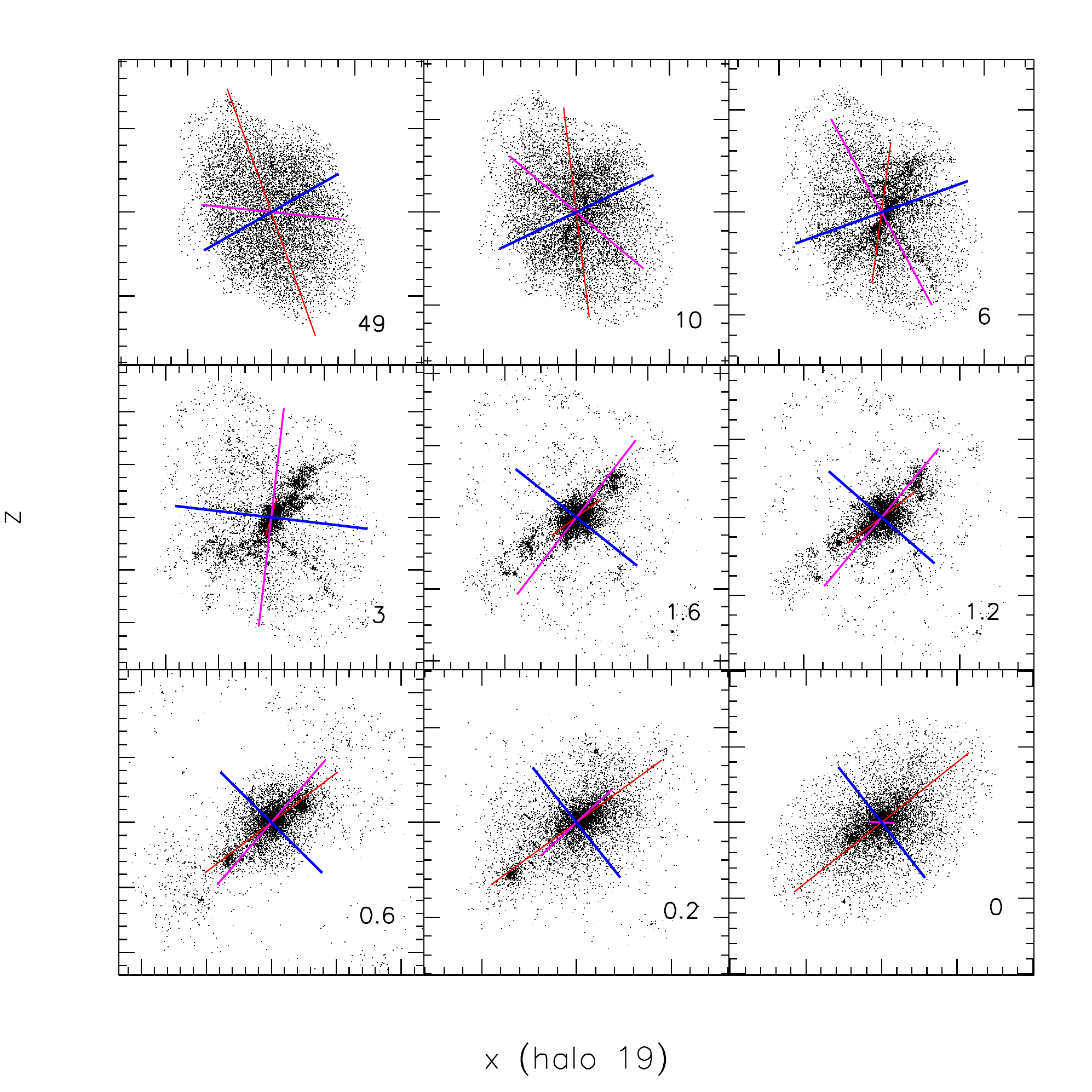}
\includegraphics[width=8cm]{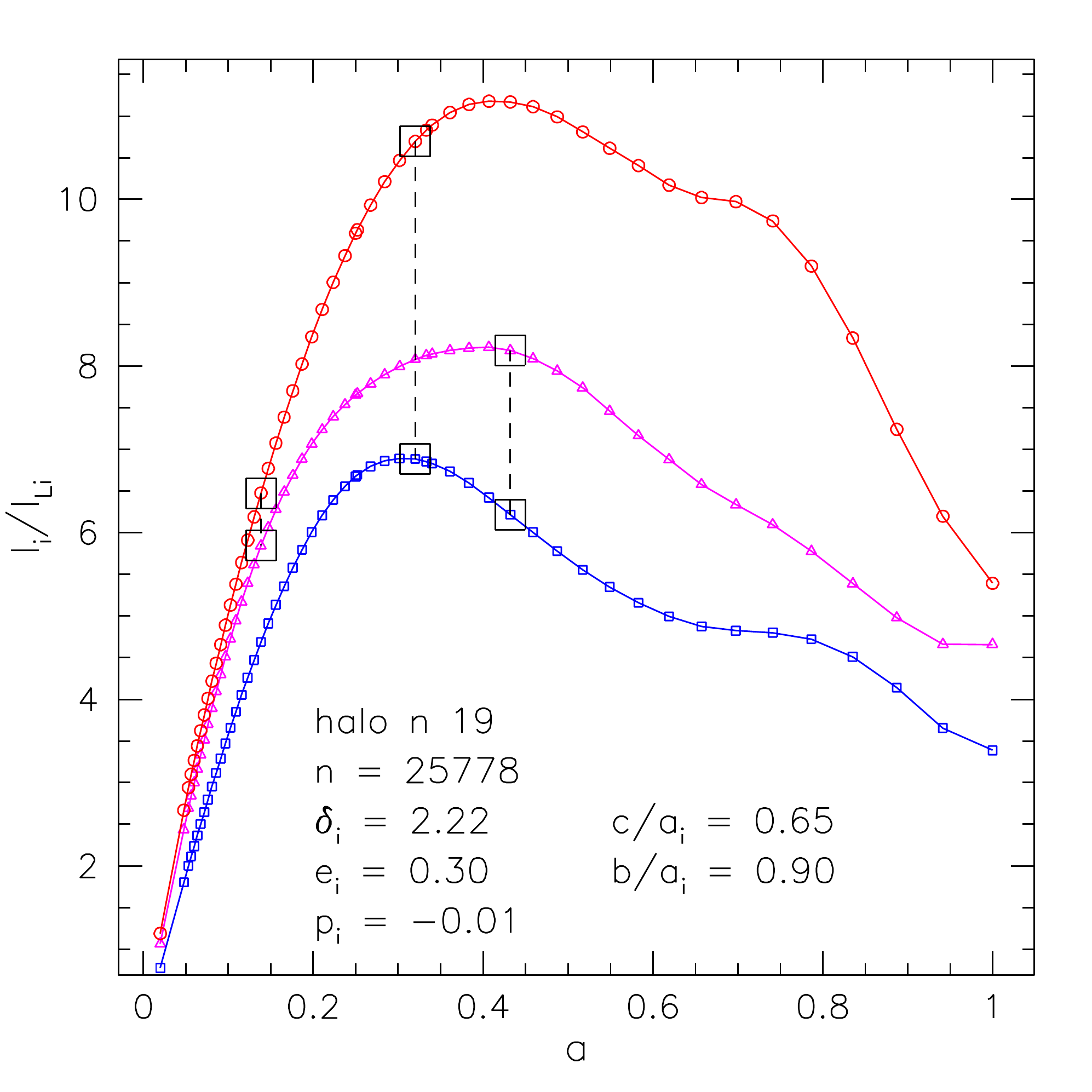}
\includegraphics[width=8cm]{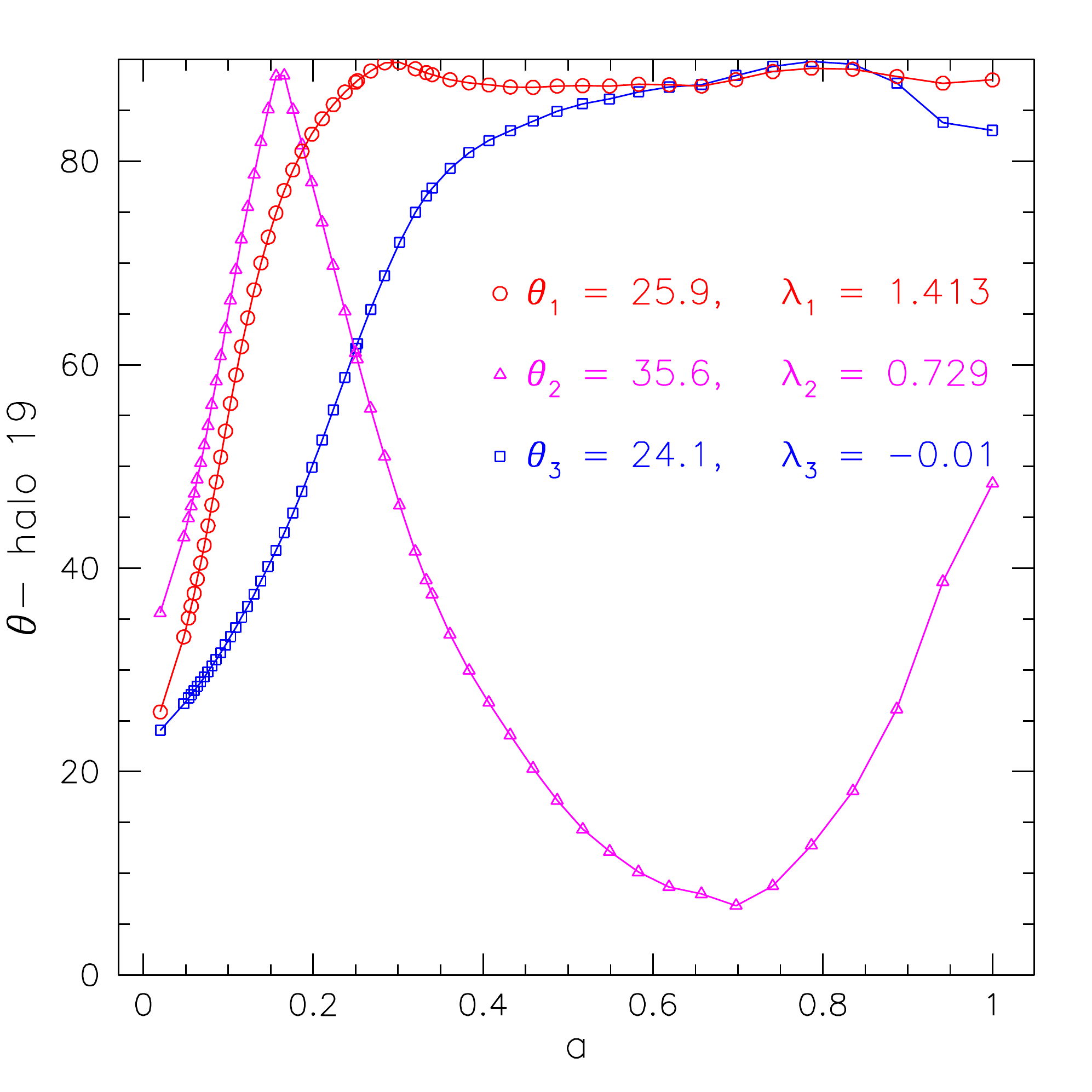}
\caption{Same as previous figure, but now for an object of mass
  $4M*$. Notice in particular that in this case $\lambda_{3}<0$.
         \label{halo4M*}}
\end{figure*}

Figure \ref{halo4M*} shows another example of evolution: the initial
parameters are again similar to the ones of the two previous haloes,
but in this case $\lambda_{3}<0$. We see that the evolution follows a
similar pattern, even if the evolution is more rapid at the beginning
and the change in direction happens more than once, involving also the
medium axis. However, at the end even this halo has a stronger
misalignment for the shortest and longest axes, while the medium
returns back towards its initial direction.

In all three cases, the misalignment angles are tens of degrees.  
This is in apparent contradiction with our finding in the main text 
that $\mu_{11}\sim 1$, which suggested perfect alignment.  However, 
note that $\cos 20^\circ = 0.93$ which is very close to unity.  

\label{lastpage}
\end{document}